\documentclass[11pt, a4paper, logo, copyright, nonumbering]{golab}
\usepackage[authoryear, sort&compress, round]{natbib}
\usepackage{dblfloatfix}
\usepackage{ulem}
\usepackage{caption}
\usepackage{dramatist}
\usepackage{xspace}
\usepackage{pifont} 
\usepackage{multirow}
\usepackage{tcolorbox}
\usepackage{xltabular}
\usepackage{longtable}
\usepackage{hyperref}
\usepackage{amsfonts}
\usepackage{amsmath}
\usepackage{amssymb}
\usepackage{lineno}
\usepackage{multirow}
\usepackage{adjustbox}
\usepackage{subcaption}

\usepackage[bottom]{footmisc}

\usepackage{CJKutf8}
\usepackage{setspace}

\usepackage{dsfont}
\usepackage{array} 
\usepackage{tabularx} 
\usepackage[labelformat=simple]{subcaption}

\usepackage{xcolor} 

\usepackage{lipsum}  
\usepackage{multicol} 
\usepackage{fontawesome5}  

\makeatletter
\def\@BTrule[#1]{%
  \ifx\longtable\undefined
    \let\@BTswitch\@BTnormal
  \else\ifx\hline\LT@hline
    \nobreak
    \let\@BTswitch\@BLTrule
  \else
     \let\@BTswitch\@BTnormal
  \fi\fi
  \global\@thisrulewidth=#1\relax
  \ifnum\@thisruleclass=\tw@\vskip\@aboverulesep\else
  \ifnum\@lastruleclass=\z@\vskip\@aboverulesep\else
  \ifnum\@lastruleclass=\@ne\vskip\doublerulesep\fi\fi\fi
  \@BTswitch}
\makeatother

\addto\extrasenglish{
}

 {\begin{list}{}%
         {\setlength{\leftmargin}{#1}}%
         \item[]%
 }
 {\end{list}}
 
\reportnumber{001} 

\newcommand{\sr}{Suiren-1.0}
\newcommand{\srb}{Suiren-Base}
\newcommand{\srd}{Suiren-Dimer}

\newcommand{\srca}{Suiren-ConfAvg}
\newcommand{\est}{EST}
\newcommand{\empp}{EMPP}
\newcommand{\qo}{Qo2mol}

\title{\centering \sr{} Technical Report: A Family of Molecular Foundation Models}

\author[*]{
Junyi An\footnote{e-mail: \textit{anjunyi@sais.org.cn}}, Xinyu Lu, Yun-Fei Shi, Li-Cheng Xu, Nannan Zhang, Chao Qu, Yuan Qi, Fenglei Cao \\
\small
Shanghai Academy of AI for Science (SAIS) \\
}

\renewcommand{\phi}{\varphi}

\renewcommand{\geq}{\geqslant}

\renewcommand{\epsilon}{\varepsilon}
\renewcommand{\imath}{\mathrm{i}}

\newlength{\restsubwidth}
\newlength{\restsubheight}
\newlength{\restsubmoreheight}
\newcommand{\rest}[2]{%
        \settowidth{\restsubwidth}{\ensuremath{#2}}
        \settoheight{\restsubheight}{\ensuremath{{}_{#2}}}
        \ensuremath{{#1\hskip 0.5pt}_{\vrule\kern2pt\parbox[b][%
        4pt][b]{\the\restsubwidth}{%
                        \ensuremath{{}_{#2}}}}}
        }

\begin{abstract}
We introduce Suiren-1.0, a family of molecular foundation models for the accurate modeling of diverse organic systems. Suiren-1.0 comprising three specialized variants (Suiren-Base, Suiren-Dimer, and Suiren-ConfAvg) is integrated within an algorithmic framework that bridges the gap between 3D conformational geometry and 2D statistical ensemble spaces. We first pre-train Suiren-Base (1.8B parameters) on a 70M-sample Density Functional Theory dataset using spatial self-supervision and SE(3)-equivariant architectures, achieving robust performance in quantum property prediction. Suiren-Dimer extends this capability through continued pre-training on 13.5M intermolecular interaction samples. To enable efficient downstream application, we propose Conformation Compression Distillation (CCD), a diffusion-based framework that distills complex 3D structural representations into 2D conformation-averaged representations. This yields the lightweight Suiren-ConfAvg, which generates high-fidelity representations from SMILES or molecular graphs. Our extensive evaluations demonstrate that Suiren-1.0 establishes state-of-the-art results across a range of tasks. All models and benchmarks are open-sourced.
\end{abstract}

\begin{document}
\begin{CJK*}{UTF8}{gbsn}

\maketitle

\begin{minipage}[t]{0.55\textwidth}  
    \vspace{0pt}  
    \section*{Model Links and Resources}
    \small
    \begin{itemize}[leftmargin=*,itemsep=1pt,parsep=0pt]
        \item[]
        \item \textbf{Suiren-Base and Suiren-Dimer Codes:} 
        \item[] \href{https://github.com/golab-ai/Suiren-Foundation-Model}{github.com/golab-ai/Suiren-Foundation-Model}
        \item[]
        \item \textbf{Suiren-ConfAvg and Finetune Codes:} 
        \item[] \href{https://github.com/golab-ai/Suiren-Property-Prediction}{github.com/golab-ai/Suiren-Property-Prediction}
        \item[]
        \item \textbf{Suiren-1.0 Model Weights:}
        \item[] \href{https://huggingface.co/ajy112/Suiren-Base}{huggingface.co/ajy112/Suiren-Base}
        \item[] \href{https://huggingface.co/ajy112/Suiren-Dimer}{huggingface.co/ajy112/Suiren-Dimer}
        \item[] \href{https://huggingface.co/ajy112/Suiren-ConfAvg}{huggingface.co/ajy112/Suiren-ConfAvg}
        \item[]
        \item \textbf{Finetune Model Weights and Agent Skills:}
        \item[] \href{https://modelscope.cn/models/ajy112/Suiren-Model-Set}{modelscope.cn/models/ajy112/Suiren-Model-Set}
        \item[] \href{https://github.com/golab-ai/Huntianling}{github.com/golab-ai/Huntianling}
        \item[] 

    \end{itemize}
\end{minipage}%
\hfill  
\begin{minipage}[t]{0.45\textwidth}  
    \vspace{0pt}  
    \centering
    \includegraphics[width=0.95\textwidth]{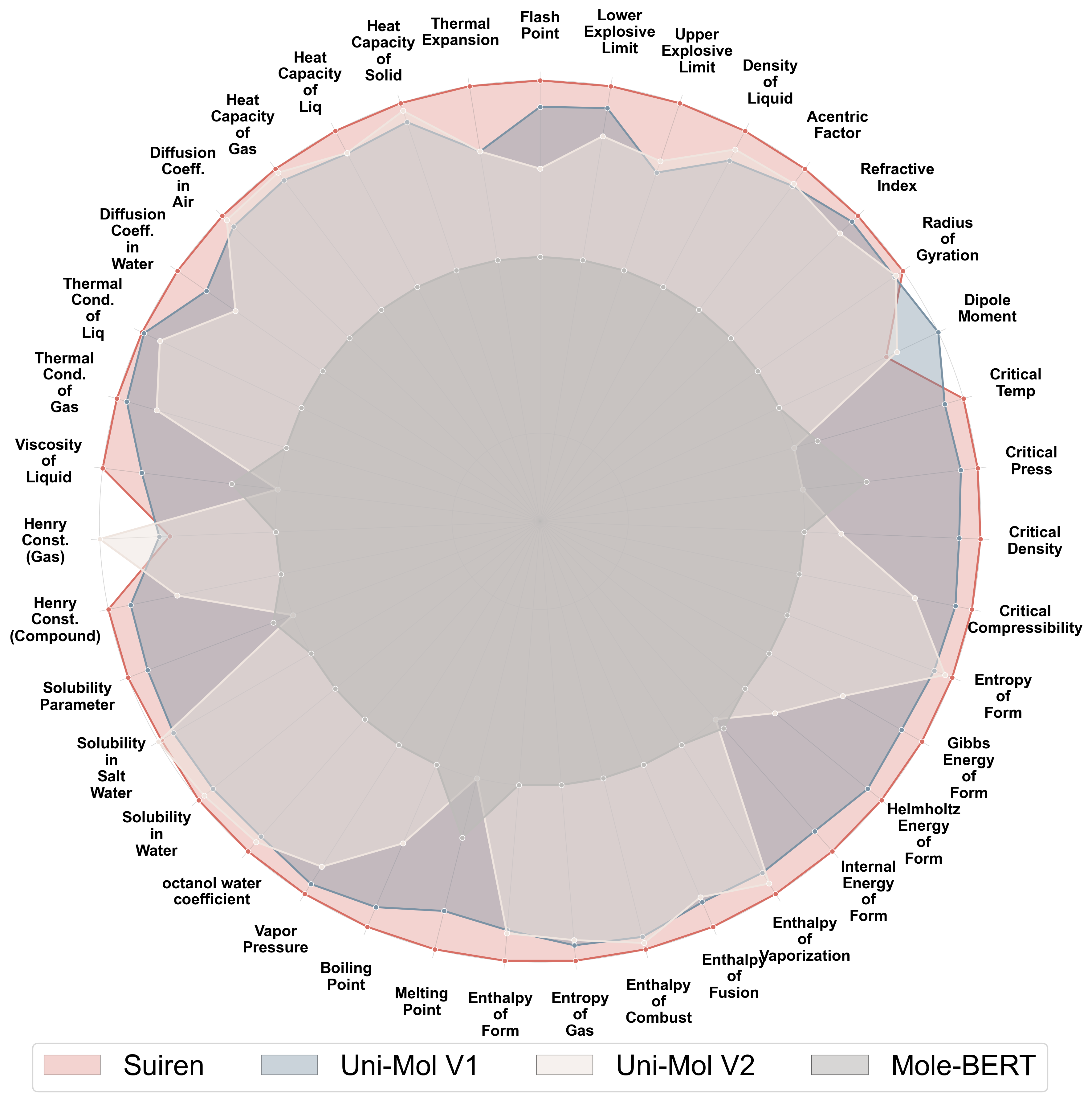}  
    \captionof{figure}{Benchmark performance of Suiren-1.0 and its counterparts. We use the normalized MAE scores ($\uparrow$).}
    \label{fig:model_effect}
\end{minipage}
\begin{figure}[htbp]
    \centering
    \begin{subfigure}[b]{0.45\textwidth}
        \includegraphics[width=\textwidth]{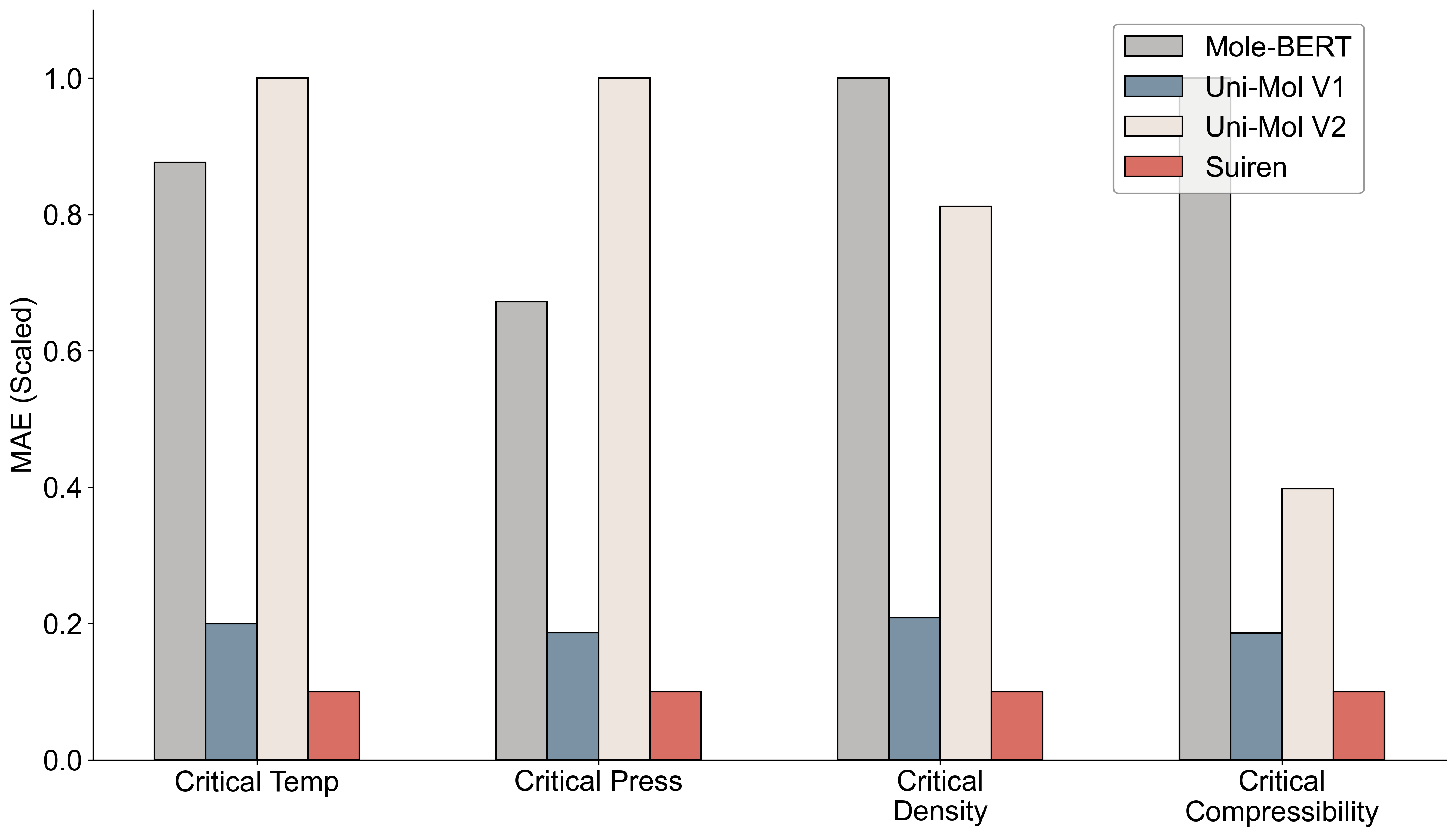}
        \caption{Critical \& Saturation Properties}
        \label{fig:critical}
    \end{subfigure}
    \hfill
    \begin{subfigure}[b]{0.45\textwidth}
        \includegraphics[width=\textwidth]{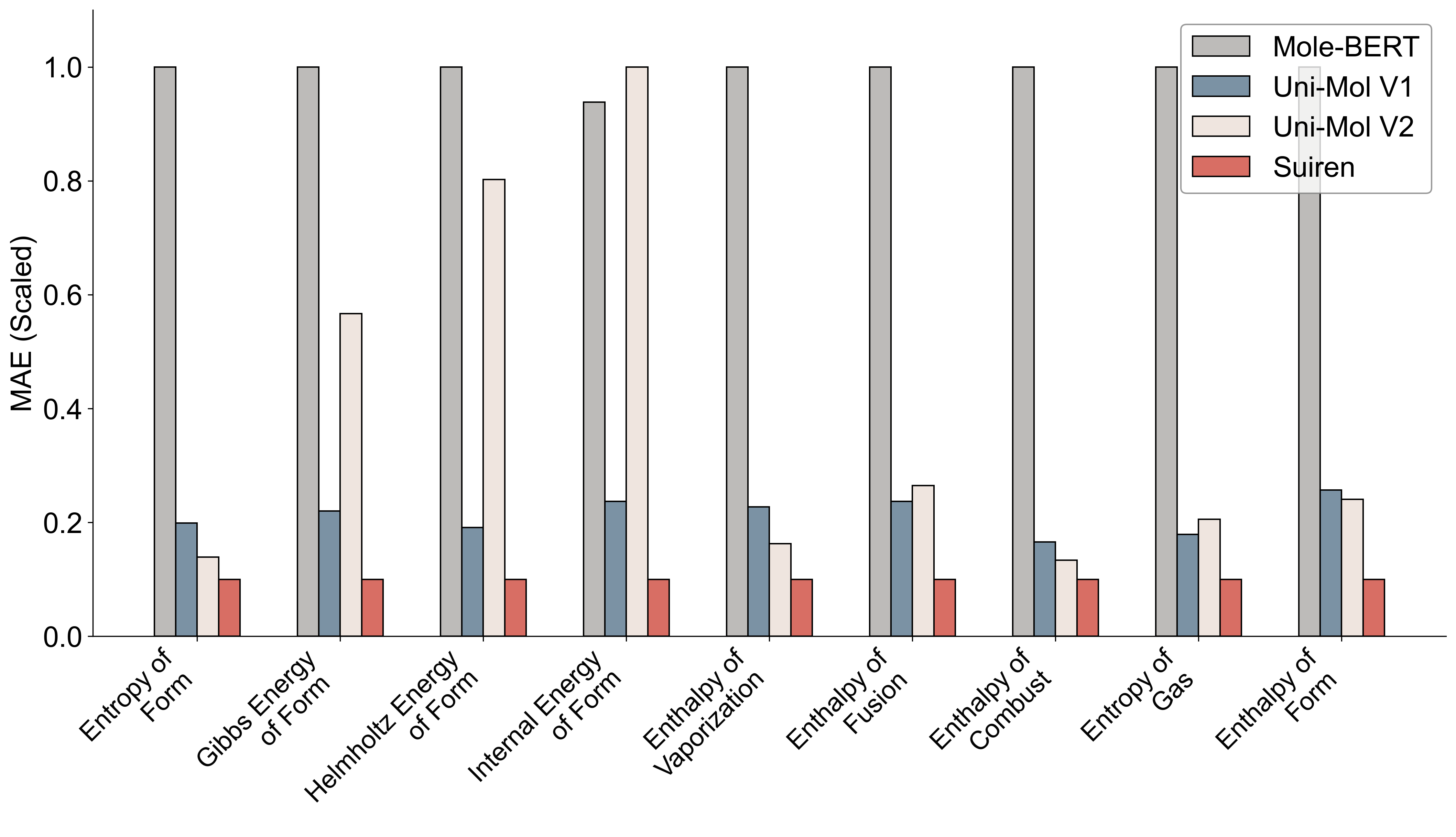}
        \caption{Energetic Properties}
        \label{fig:energetic}
    \end{subfigure}
    
    \begin{subfigure}[b]{0.45\textwidth}
        \includegraphics[width=\textwidth]{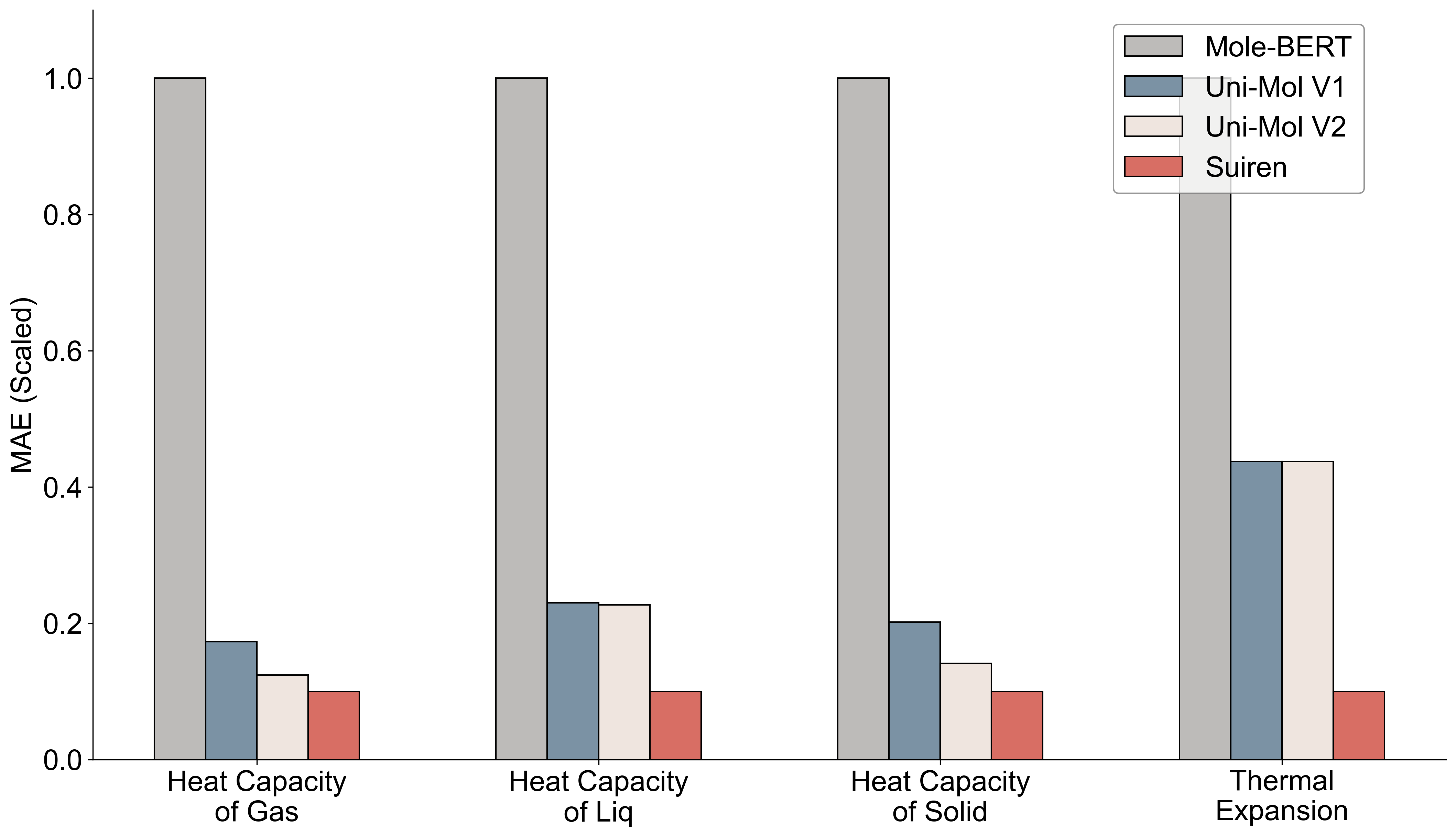}
        \caption{Fluctuation Properties}
        \label{fig:fluctuation}
    \end{subfigure}
    \hfill
    \begin{subfigure}[b]{0.45\textwidth}
        \includegraphics[width=\textwidth]{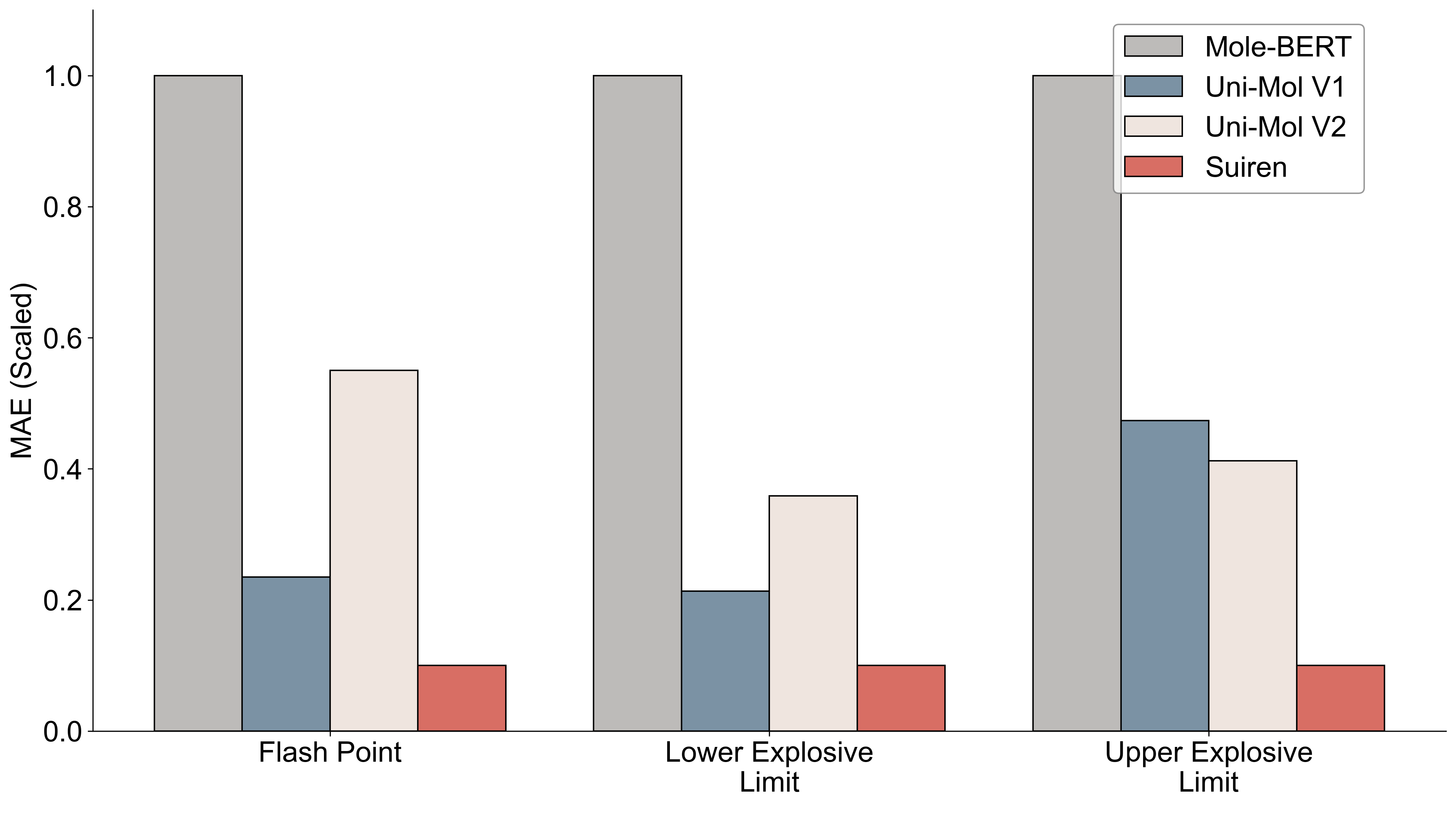}
        \caption{Safety Properties}
        \label{fig:safety}
    \end{subfigure}
    
    \begin{subfigure}[b]{0.45\textwidth}
        \includegraphics[width=\textwidth]{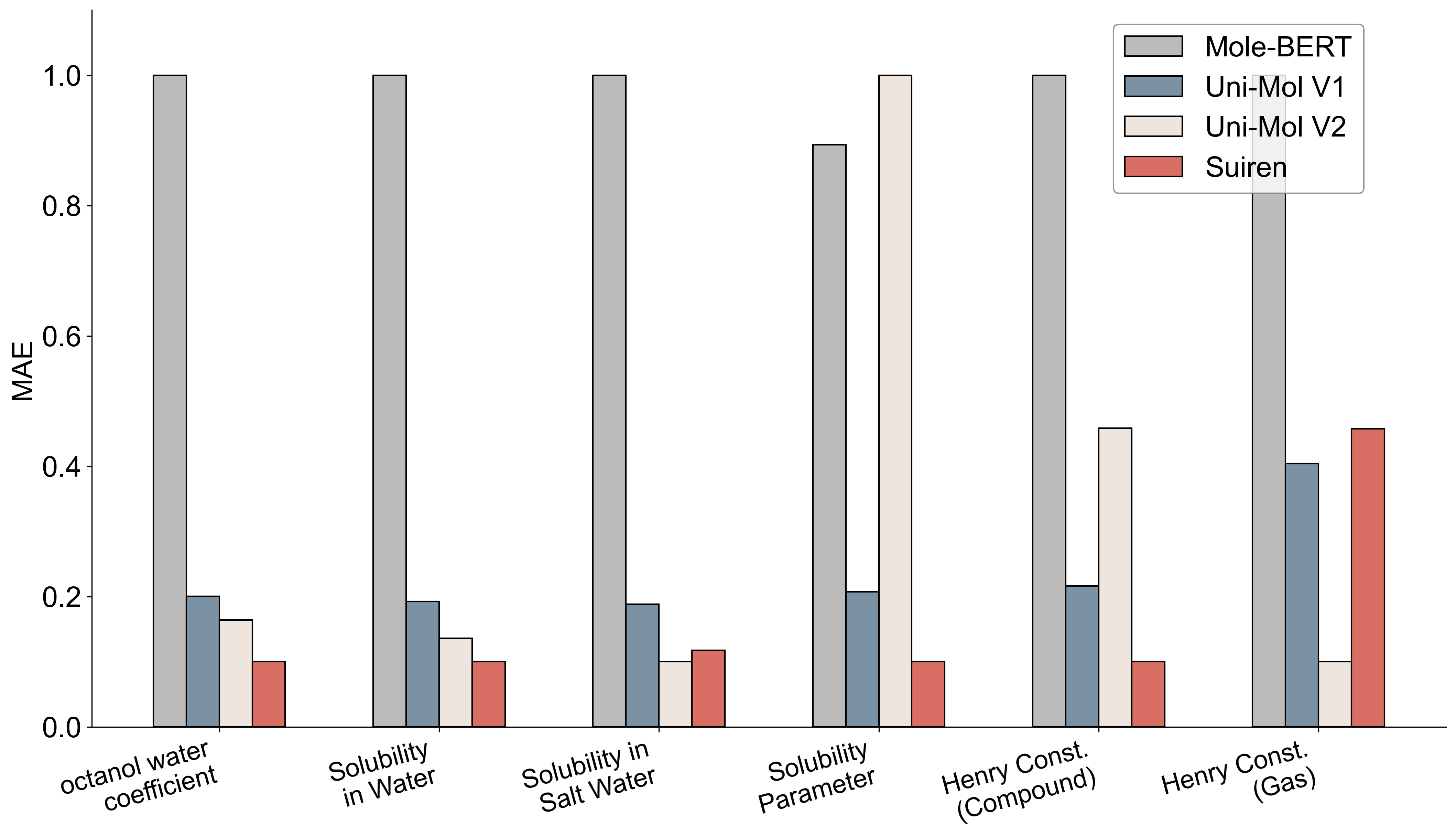}
        \caption{Solution Properties}
        \label{fig:solution}
    \end{subfigure}
    \hfill
    \begin{subfigure}[b]{0.45\textwidth}
        \includegraphics[width=\textwidth]{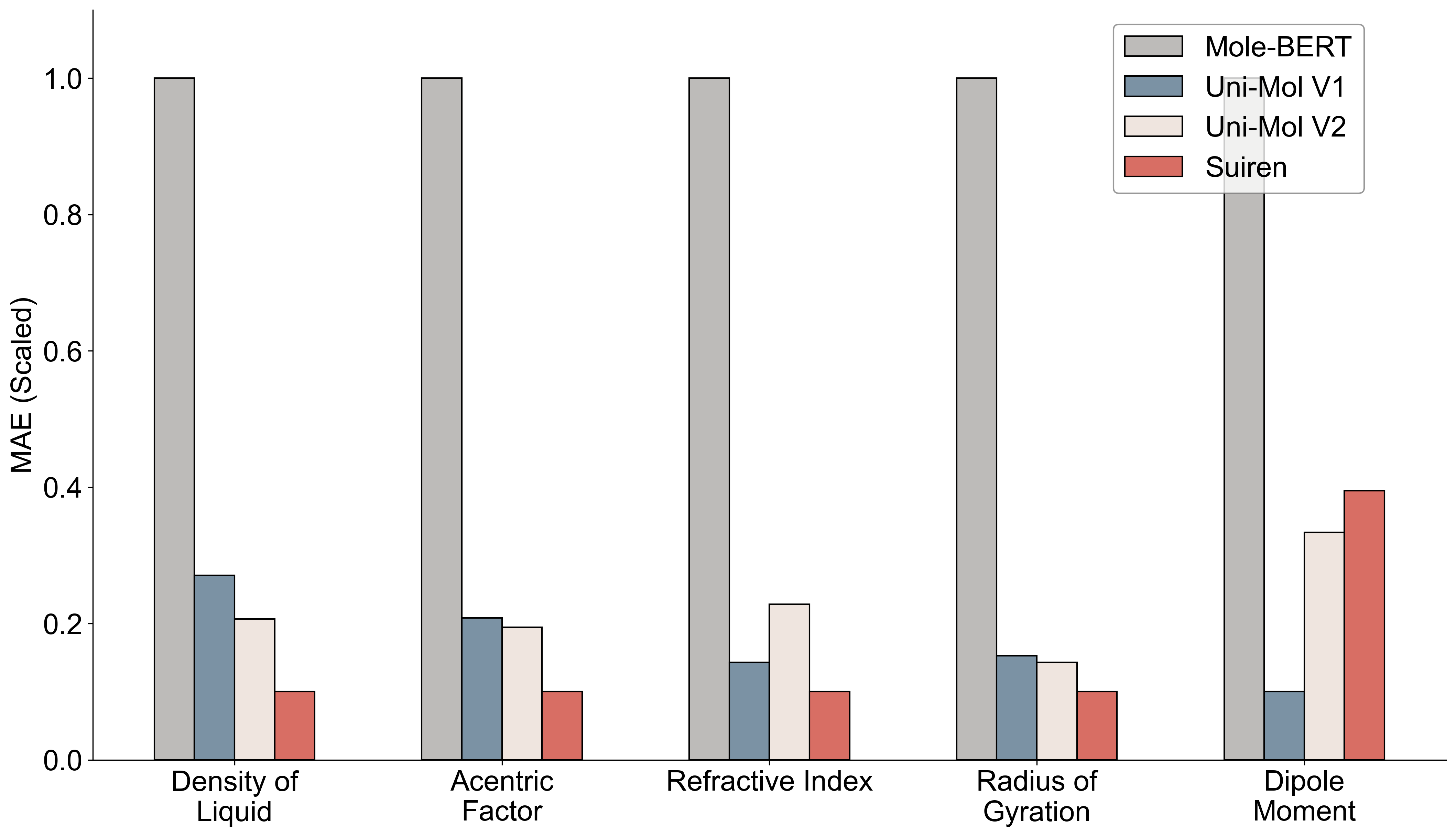}
        \caption{Structural Properties}
        \label{fig:structural}
    \end{subfigure}
    
    \begin{subfigure}[b]{0.45\textwidth}
        \includegraphics[width=\textwidth]{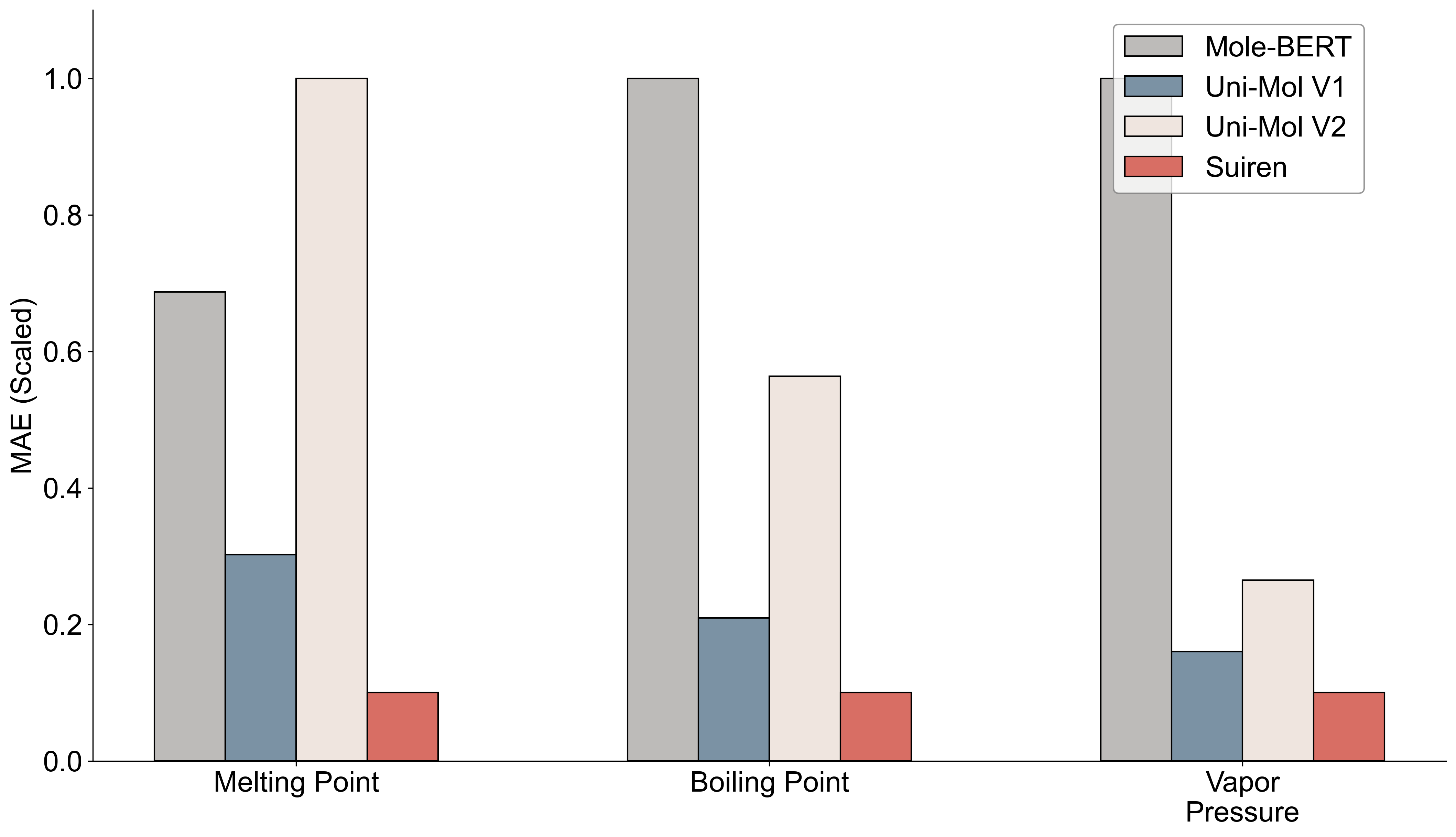}
        \caption{Thermal Properties}
        \label{fig:thermal}
    \end{subfigure}
    \hfill
    \begin{subfigure}[b]{0.45\textwidth}
        \includegraphics[width=\textwidth]{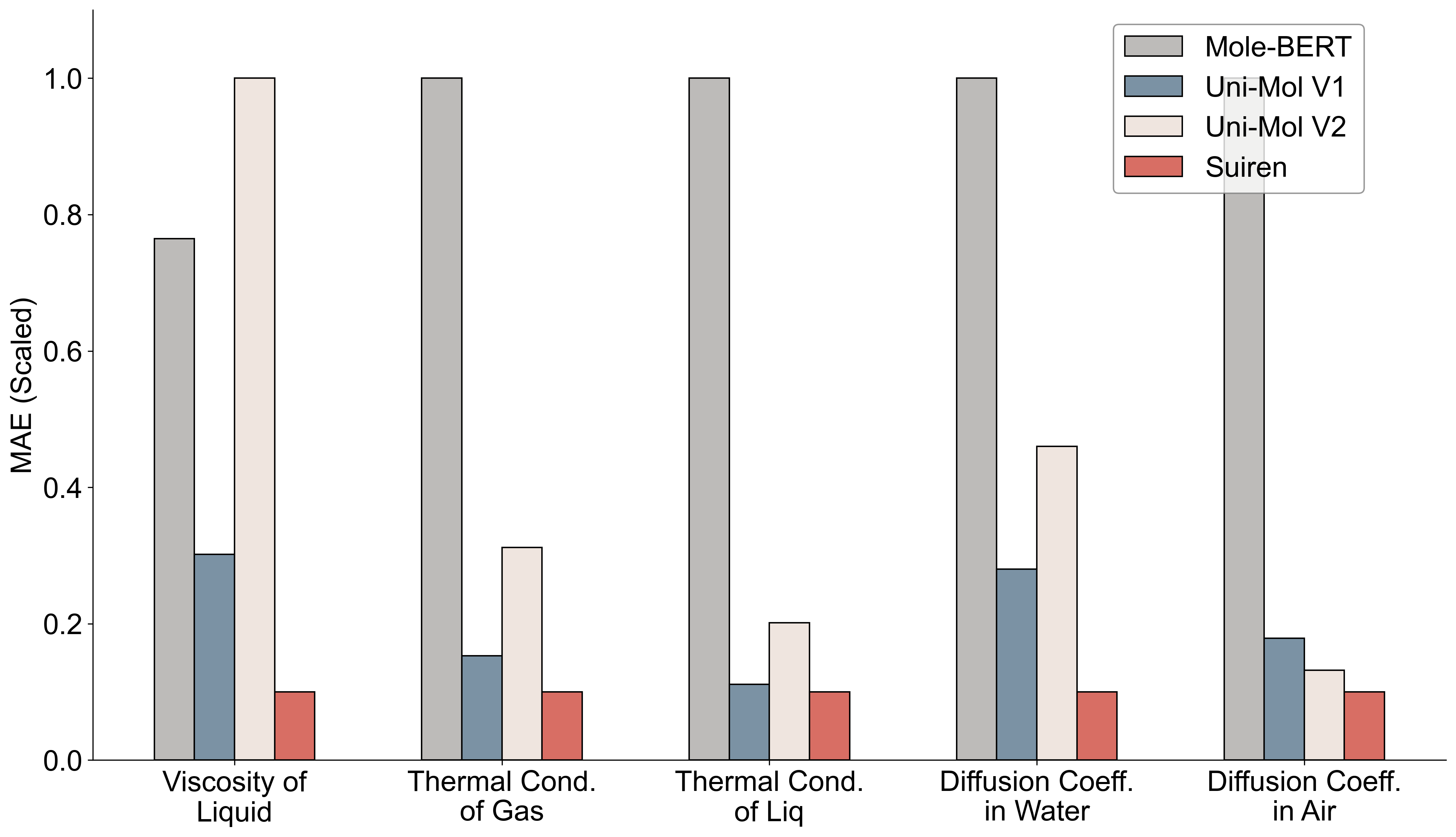}
        \caption{Transport Properties}
        \label{fig:transport}
    \end{subfigure}
    
    \caption{Comparison of Suiren-1.0 model and molecular Foundation model across various tasks in 8 domains. All tasks are regression tasks, with MAE ($\downarrow$) as the evaluation metric. Due to significant differences in metric ranges across different tasks, the y-axis is scaled.}
    \label{fig:4x2property_prediction}
\end{figure}

\newpage

\section{Introduction}

\begin{figure}
    \centering
    \includegraphics[width=1.0\linewidth]{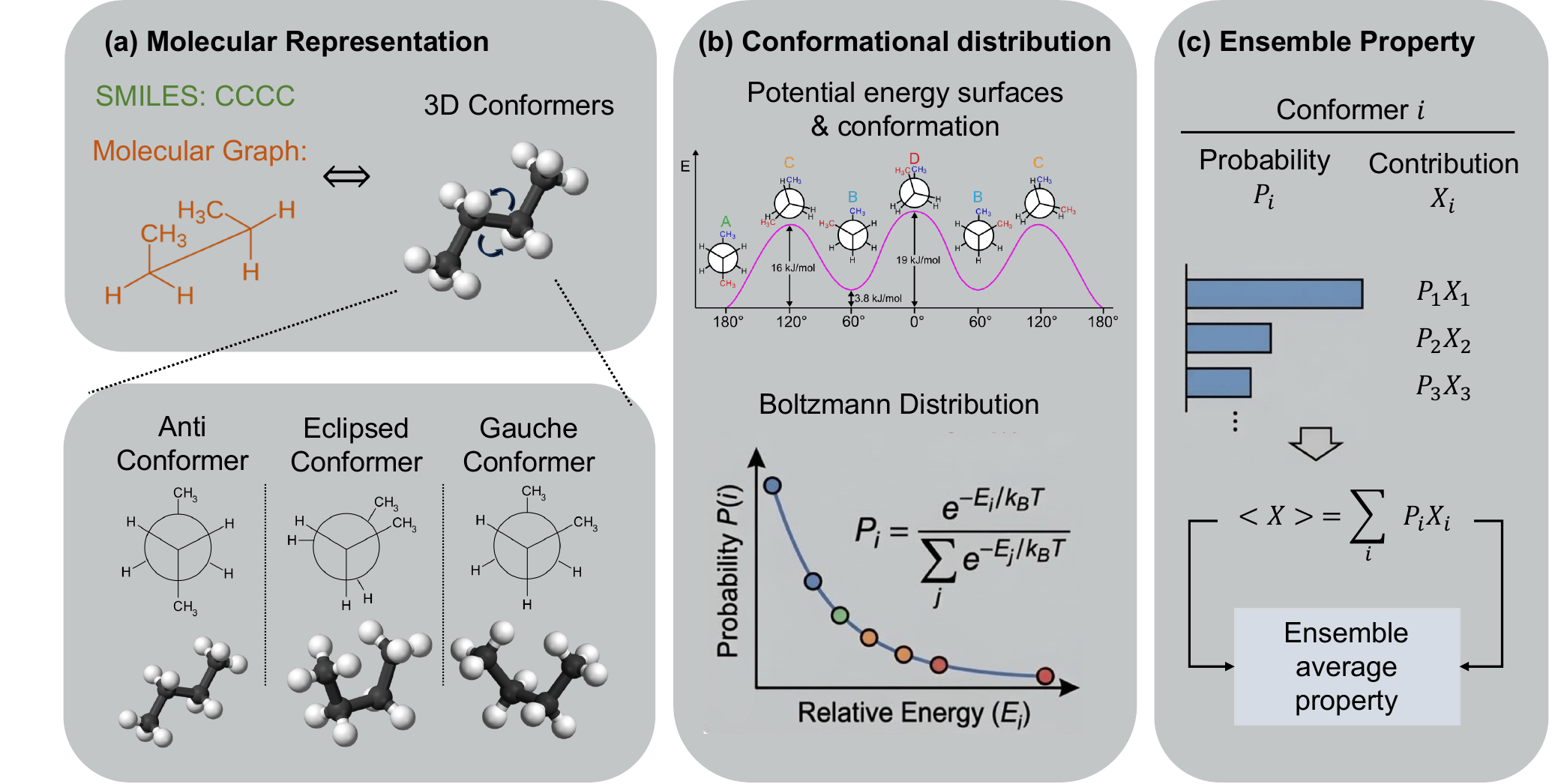}
    \caption{\textbf{Microscopic and macroscopic representations of molecular ensembles.} \textbf{(a) Molecular Representation:} A single molecular identity corresponds to a diverse ensemble of 3D conformations at the microscopic space. \textbf{(b) Conformational distribution:} The relative probability of these conformations is governed by the Boltzmann distribution as a function of potential energy. \textbf{(c) Ensemble Property:} Macroscopic observables emerge as the ensemble-averaged properties derived from the collective contributions of all constituent conformations.}
    \label{fig:boltzmann}
\end{figure}

Foundation models have catalyzed a paradigm shift in natural language processing and computer vision, where large-scale pre-training facilitates robust transferability across diverse downstream tasks \citep{achiam2023gpt, team2023gemini, yang2025qwen3, liu2024deepseek}. In the science domain, pioneering molecular architectures such as MoleBERT \citep{xia2023mole}, Uni-Mol \citep{zhou2023uni, ji2024uni}, and UMA \citep{wood2025family} have demonstrated significant promise. However, compared to the linguistic and visual domains, universal molecular modeling remains hindered by inherent scientific complexities and a scarcity of high-quality supervised data. We identify the primary challenges as follows:

\begin{itemize}
\item First, the "physical priors" governing molecular systems are exceptionally complex. Molecular behavior is dictated by the intricate laws, such as quantum mechanics (e.g., the Schrödinger equation) and statistical thermodynamics (e.g., Boltzmann distributions) \citep{schleich2013schrodinger, charbonneau1982linear}. Capturing these fundamental mechanisms solely through data-driven learning is challenging, particularly given the sparsity of high-fidelity labeled data.

\item Second, a persistent multiscale gap remains between microscopic structures and macroscopic observables. Microscopic tasks typically demand the resolution of explicit 3D conformations and electronic densities, where Density Functional Theory (DFT) enables the generation of abundant, high-quality labeled data \citep{liu2024open,chanussot2021open,levine2025open}. Conversely, macroscopic tasks often rely on 1D SMILES or 2D molecular graphs that lack explicit conformational information. While these tasks span broad chemical spaces, their data is often scarce, as macroscopic labels frequently require costly wet-lab experiments or molecular dynamics simulations. Physically, these two modalities are intrinsically linked: macroscopic features emerge from the ensemble-averaged properties of a molecule’s conformations, governed by the Boltzmann distribution (see Figure \ref{fig:boltzmann}). However, existing approaches largely fail to bridge this divide. Pure 3D foundation models, such as UMA, learn rich 3D representations from labeled data but lack generalizability across broad chemical tasks; meanwhile, pure 2D models, such as Mole-BERT, capture graph topology through self-supervised learning yet remain "conformation-blind," limiting their predictive effectiveness.
\end{itemize}

\noindent \sr{} is designed to bridge the multiscale gap between microscopic and macroscopic representations. We first pre-train \srb{} (1.8B parameters) on large-scale, first-principles quantum chemical data using objectives specifically tailored for microscopic, conformation-aware representation learning. We then introduce Conformation Compression Distillation (CCD), a diffusion-based strategy that distills the knowledge of \srb{} into \srca{}. This process encodes a macroscopic latent representation that can be inverted into specific 3D conformations through energy-conditioned queries. In the absence of an energy query, \srca{} accepts 2D molecular graphs or 1D SMILES as input to produce generalizable molecular embeddings suitable for a wide range of downstream tasks, including materials discovery, drug design, and battery chemistry.

\noindent We evaluate Suiren-1.0 across a comprehensive suite of over 50+ tasks spanning 9 diverse scientific domains. To ensure a rigorous and fair assessment, we eschew task-specific engineering in favor of a unified fine-tuning and inference pipeline across all benchmarks. As illustrated in Figure \ref{fig:model_effect} and Figure \ref{fig:4x2property_prediction}, Suiren-1.0 achieves consistent, state-of-the-art (SOTA) performance in the vast majority of cases. Notably, Suiren-1.0 delivers performance gains exceeding 20\% on more than 20+ tasks compared to existing models. We attribute this success to the synergy between large-scale model scaling and the principled integration of physical priors. 

\noindent Our main contributions are as follows:

\noindent
\textbf{Modeling Framework: Microscopic--Macroscopic Bridging}
\begin{itemize}[topsep=0pt]
    \item
    We establish a three-stage framework to unify molecular scales: (i) pre-training a 3D conformation-aware foundation model for high-fidelity microscopic representation learning; (ii) distilling this knowledge into a compressed, conformation-agnostic model for macroscopic adaptation via Conformation Compression Distillation; and (iii) fine-tuning task-specific encoders for a diverse suite of downstream scientific applications.
\end{itemize}

\noindent
\textbf{Pre-Training: Physical Priors and First-Principles Data}
\begin{itemize}[topsep=0pt]
    \item
    We train \srb{} on large-scale first-principles quantum-chemical data (\qo{} \cite{liu2024open}) and incorporate physically motivated algorithms, including advanced \empp{} \citep{an2025empp} and advanced \est{} \citep{an2025est}, to improve representation quality.
    \item
    The resulting representations capture conformation-sensitive microscopic information and transfer effectively across downstream scientific tasks.
    We further provide a continued pre-training variant for dimer systems (Suiren-Dimer).
\end{itemize}

\noindent
\textbf{Transfer Learning: Broad Molecular Applicability}
\begin{itemize}[topsep=0pt]
    \item
    By distilling from \srb{} to \srca{}, we enable strong performance when only graph or SMILES inputs are available, improving deployability in real-world molecular pipelines.
    \item
    We benchmark \srca{} on 50+ property prediction tasks and observe robust improvements over advanced molecular baselines.
\end{itemize}
\noindent
\textbf{Open Science}
\begin{itemize}[topsep=0pt]
    \item
    We provide a description of pre-training, distillation, and finetune models, and release their weights to facilitate reproducible molecular foundation-model research. Furthermore, we open a comprehensive benchmark MoleHB for molecular model evaluation.
\end{itemize}

\noindent The remainder of this paper is organized as follows. Section~\ref{sec:arch} details the architectural design and the broader \sr{} model family. Section~\ref{sec:pre-training} describes our data curation and pre-training methodology, followed by Section~\ref{sec:post-training}, which introduces our post-training distillation and fine-tuning protocols. In Section~\ref{sec:evaluation}, we present a comprehensive evaluation of Suiren-1.0. Finally, Section~\ref{sec:conclusion} concludes the paper with a discussion on current limitations and future research directions.
\section{Architecture}
\label{sec:arch}

\subsection{Large SO(3)-Equivariant Graph Neural Network}

\srb{} is a high-degree equivariant graph neural network (GNN) designed for 3D conformational representation learning. The architecture integrates an EquiformerV2 model \citep{liao2023equiformerv2} with a dense Mixture-of-Experts (MoE) update block, which concurrently utilizes both S2Activation and Equivariant Spherical Transformer (EST) experts \citep{an2025est} in each forward pass. Following the standard message-passing framework \citep{gilmer2017neural}, the model architecture is formulated as:
\begin{align}\mathbf{m}_{i}^{(k)} &= \sum_{j \in \mathcal{N}(i)} \psi_{m}^{(l)}\left(\mathbf{x}_{i}^{(l)}, \mathbf{x}_{j}^{(l)}, \mathbf{e}_{ij}\right), \\
\mathbf{x}_{i}^{(k+1)} &= \psi_{u}^{(l)}\left(\mathbf{x}_{i}^{(l)}, \mathbf{m}_{i}^{(l)}\right),
\end{align}
where $N(i)$ denotes the neighbor set of node $i$, $\mathbf{x}_i$ represents the node embeddings, $\mathbf{e}_{ij}$ denotes the edge features, and $\psi_{m}^{(l)}$ and $\psi_{u}^{(l)}$ correspond to the message and update functions, respectively. The message block captures interatomic interactions with a computational complexity linear in the number of edges, and aggregates these into node-level messages. Conversely, the update block processes these messages with a complexity linear in the number of nodes. These components serve functional roles analogous to the self-attention and feed-forward network (FFN) modules in standard Transformer architectures.

\begin{figure}[t]
    \centering
    \includegraphics[width=1.0\linewidth]{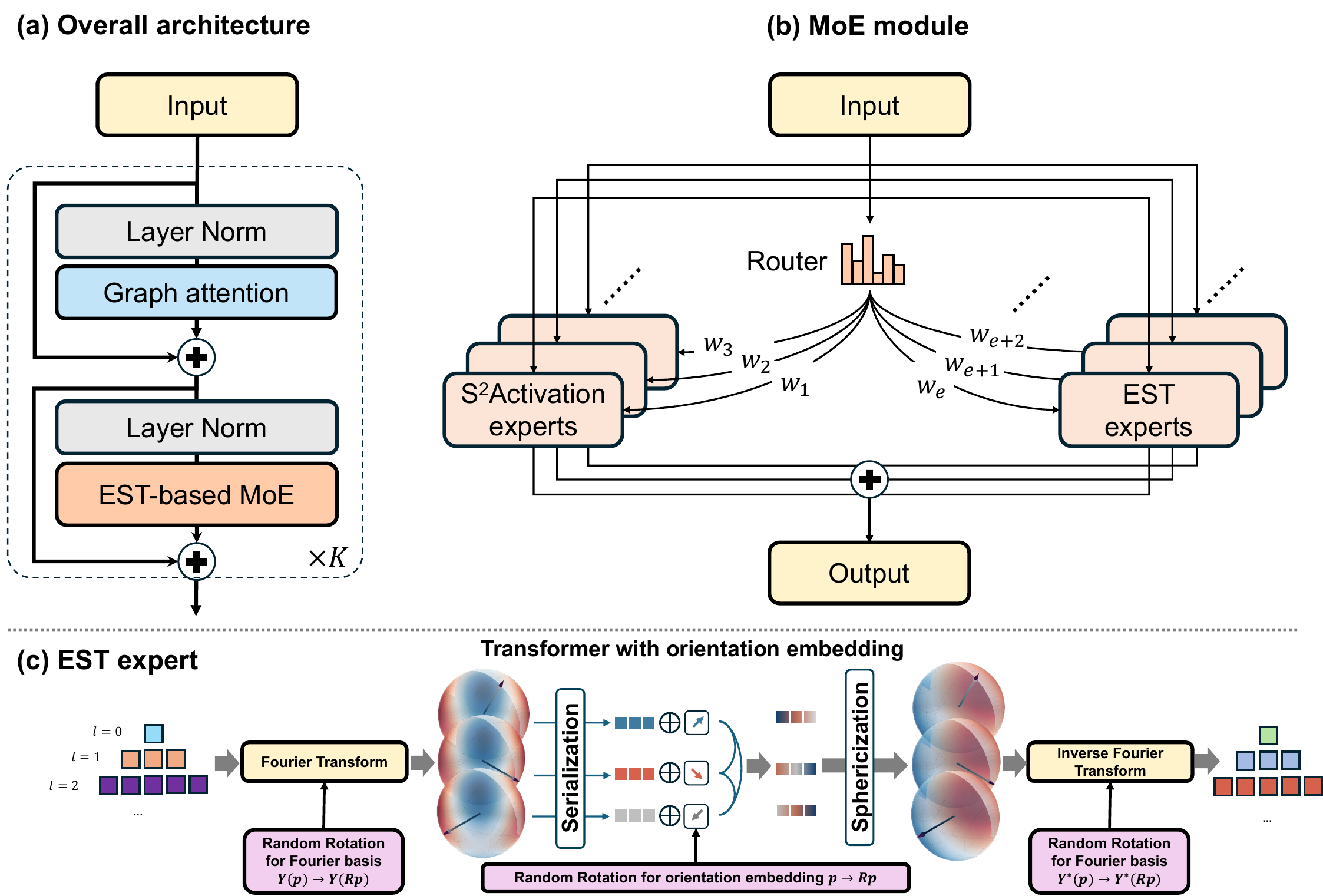}
    \caption{\textbf{The architecture of the Suiren-Base model. (a) Overall framework. (b) A dense MoE block. (c) Modified EST expert:} during training, the spherical Fourier transform basis set and orientation embedding are subjected to a random rotation.}
    \label{fig:3Dmodel}
\end{figure}

\noindent Figure \ref{fig:3Dmodel} illustrates the architecture of \srb{}. The message block is adapted from the EquiformerV2 graph attention block and utilizes an $SO(2)$-linear operation to integrate edge features $\mathbf{e}_{ij}$ with node attributes $(\mathbf{x}_{i}, \mathbf{x}_{j})$. For the update block, we employ a dense MoE design to enhance model capacity. This architectural choice is driven by two primary observations: (1) the update block is computationally efficient, allowing the addition of experts with minimal latency overhead; and (2) the complexity of the aggregated messages benefits significantly from increased parameter capacity.

\noindent \srb{} contains 20 layers ($K=20$), and each MoE block contains 20 S2Activation experts and 20 EST experts, balancing equivariance and expressiveness.
The original EST maps steerable group embeddings from the harmonic domain to spatial domain with Fourier basis sampled spherical points, updates embeddings with a spherical Transformer, and projects them back:
\begin{small}
\begin{align}
    f(\vec{\mathbf{p}})&=\mathcal{F}(\mathbf{x})=\sum_{l=0}^{\infty}\sum_{m=-l}^{l}\mathbf{x}^{(l,m)}Y^{(l,m)}(\vec{\mathbf{p}}) &\text{\# Fourier transform on a single spherical point} \\
    \mathbf{f} &= [f(\vec{\mathbf{p}}_1), f(\vec{\mathbf{p}}_2), ..., f(\vec{\mathbf{p}}_S)] &\text{\# The spatial representation on sampling points $\vec{\mathbf{p}}_i$} \\
    \hat{\mathbf{f}} &= \mathrm{Trans}([\mathbf{f};\mathbf{P}]) &\text{\# Transformer with orientation embedding $\mathbf{P}=[\vec{\mathbf{p}}_1,...,\vec{\mathbf{p}}_S]$} \\
    \hat{\mathbf{x}} &= \sum_{s=1}^{S}  \hat{\mathbf{f}}_s \cdot \mathbf{Y}^{*}(\vec{\mathbf{p_s}})&\text{\# Back to harmonic domain}
\end{align}
\end{small}

\noindent where $Y^{l,m}(\cdot)$ and $\mathbf{Y}(\cdot)=[Y^{l_1,m_1}(\cdot),Y^{l_2,m_2}(\cdot), ...]$ denote the spherical harmonic basis (scale and vector), $\vec{\mathbf{p}}$ denotes the sample point (orientation) in the sphere, and the term $\hat{\cdot}$ is used to represent the embedding update. Although uniform spherical sampling in EST offers partial equivariance, it remains susceptible to discretization-induced errors. To mitigate these artifacts, we propose a basis-rotation strategy for EST experts. During training, the Fourier basis for each sample in a mini-batch is pre-rotated by a random 3D rotation $\mathbf{R}$. Since this procedure only modifies the basis orientation, the computational overhead remains negligible. This approach exposes the model to a diverse range of orientations, encouraging the learning of orientation-consistent responses and more closely approximating continuous spherical Fourier behavior. The formulation is defined as:
\begin{align}
    \mathbf{f} &= [f(\mathbf{R}\vec{\mathbf{p}}_1), f(\mathbf{R}\vec{\mathbf{p}}_2), ..., f(\mathbf{R}\vec{\mathbf{p}}_S)]  \\
    \hat{\mathbf{f}} &= \mathrm{Trans}([\mathbf{f};\mathbf{R}\mathbf{P}]) \\
    \hat{\mathbf{x}} &= \sum_{s=1}^{S}  \hat{\mathbf{f}}_s \cdot \mathbf{Y}^{*}(\mathbf{R}\vec{\mathbf{p_s}}).
\end{align}
By leveraging this adaptive equivariance mechanism, the spherical sampling density $S$ can be optimized toward the Nyquist-rate lower bound, $S \geq (2L)^2$, where $L$ denotes the maximum degree of the spherical harmonic embedding. This reduction significantly lowers both training and inference overhead without compromising the robustness of the model's equivariant properties.

\noindent Both \srb{} and \srd{} utilize a unified backbone architecture. In practice, these models take 3D atomic coordinates of molecules or dimers as input to predict quantum-accurate (DFT-level) potential energies and interatomic forces.

\subsection{Conformation Compression Distillation}\label{sec:CCD}
As illustrated in Figure \ref{fig:boltzmann}, molecular properties are typically determined by ensemble-averaged behavior across multiple physically plausible conformers. While these conformer probabilities are governed by the Boltzmann distribution, the exact distribution—and the underlying potential energy surface (PES)—is generally unknown a priori. To address this, we propose Conformation Compression Distillation (CCD), a feature-alignment framework designed for one-to-many molecule-conformer mapping.

\begin{figure}[htbp]
    \centering
    \includegraphics[width=0.85\linewidth]{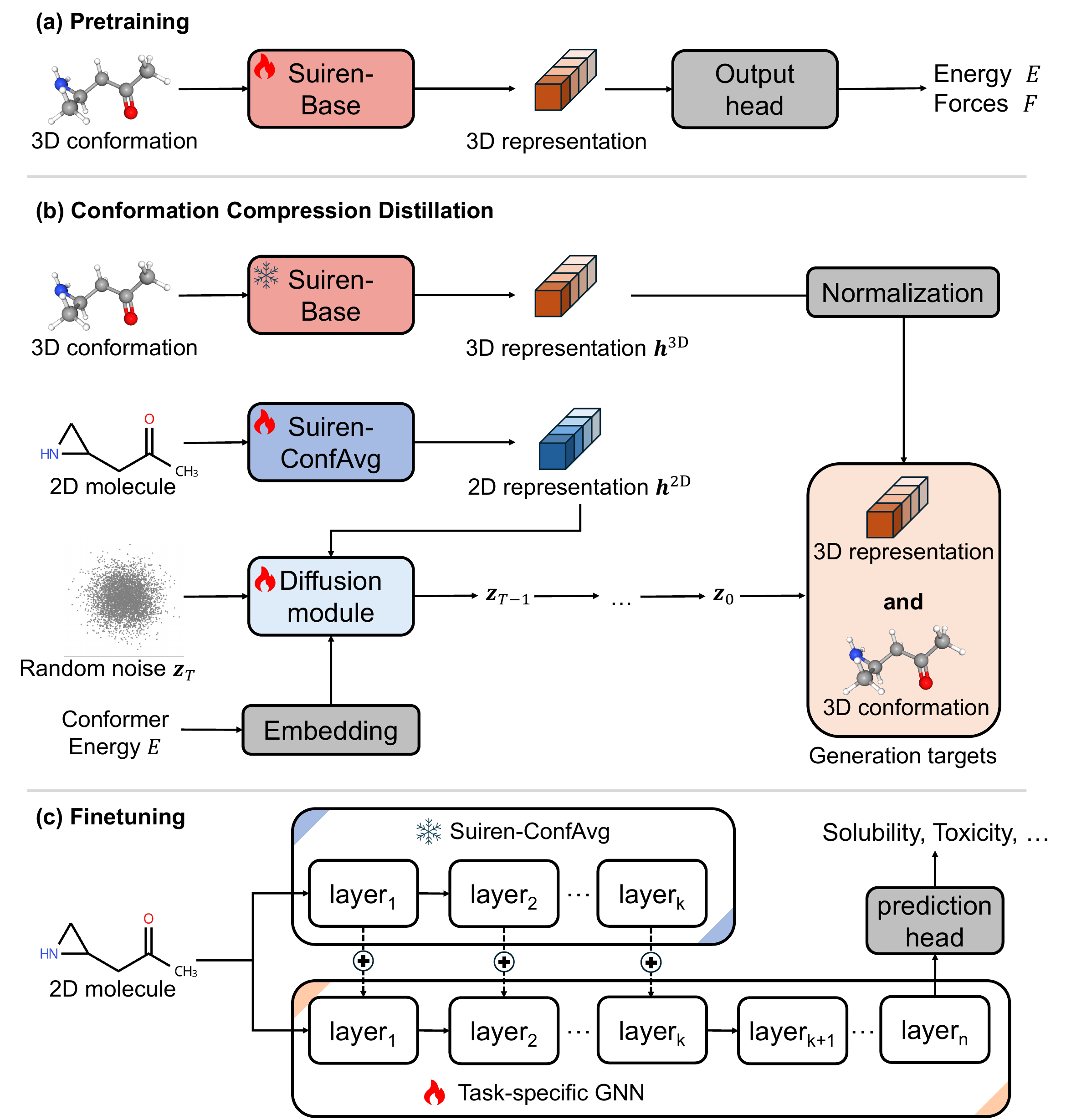}
    \caption{\textbf{Overview of training stages.} \textbf{(a) 3D Pre-training:} Self-supervised learning on 3D molecular conformations. \textbf{(b) Conformation Distillation:} Distilling 3D geometric knowledge into a conformation-averaged representation. \textbf{(c) Downstream Fine-tuning:} Adapting the model for supervised molecular property prediction.}
    \label{fig:training_stage}
\end{figure}

\noindent For each molecule-conformer pair, CCD operates on two distinct modalities: the 2D molecular topology (SMILES or graph) and the 3D conformer with its associated energy $E$. The 2D input is processed via a Graph Attention Network (GAT) to extract a latent representation $\mathbf{h}^{\mathrm{2D}}$, while the 3D conformer is encoded by a pre-trained \srb{} teacher to yield an equivariant representation $\mathbf{h}^{\mathrm{3D}}$. We then introduce a 3D diffusion-based model featuring a lightweight Equiformer+MoE+EST dynamics network $\phi_{\theta}(\cdot)$. This network is conditioned on both $\mathbf{h}^{\mathrm{2D}}$ and the energy $E$, with the diffusion process targeting the joint reconstruction of the 3D representation $\mathbf{h}^{\mathrm{3D}}$ and the molecular 3D coordinates $\mathbf{c}$ (see Figure \ref{fig:training_stage}(b)).

\noindent During training, we add random noise to the clean 3D target state:
\begin{equation}
    \mathbf{z}_t = \alpha_t\mathbf{z}_{-1} + \sigma_t \boldsymbol{\epsilon}
\end{equation}
where $\mathbf{z}_{-1}=[\mathbf{h}^{\mathrm{3D}}; \mathbf{r}]$ denotes the clean target state, $\mathbf{z}_t$ is the noisy state at timestep $t\in [0, T]$, $\alpha_t,\sigma_t$ are schedule coefficients, and $\boldsymbol{\epsilon}$ denotes the Gaussian noise.
We freeze the weights of \srb{}, and train the dynamics network and 2D representation model by predicting the noise:
\begin{align}    
    \hat{\boldsymbol{\epsilon}} &= \phi_{\theta}(\mathbf{z}_t, t, \mathbf{h}^{\mathrm{2D}}, E)  \\
    \mathcal{L}_{\mathrm{CCD}} &= \operatorname{MSE}\!\left(\hat{\boldsymbol{\epsilon}}, \boldsymbol{\epsilon}\right),
\end{align}
where $E$ is encoded with Gaussian embeddings and $\mathcal{L}_{\mathrm{CCD}}$ denote optimize objective functions.

\noindent Upon completion of training, the framework yields two primary components: a 2D molecular encoder and a generative diffusion dynamics model. These modules facilitate both high-fidelity conformer generation and robust 2D representation learning. For conformer generation, the diffusion dynamics model captures the multimodal nature of the structural space, generating diverse ensembles rather than collapsing to a single low-energy mode \citep{Landrum2016RDKit2016_09_4,xu2024gtmgc}. The reverse-time sampling step from timestep $t$ to $s = t - 1$ is formulated as:
\begin{equation}
    \mathbf{z}_s = \frac{1}{\alpha_{t|s}}\mathbf{z}_t - \frac{\sigma_{t|s}^2}{\alpha_{t|s}\sigma_t}\cdot \phi_{\theta}(\mathbf{z}_t, t, \mathbf{h}^{\mathrm{2D}}, E) + \sigma_{t\to s}\cdot \boldsymbol{\epsilon},
\end{equation}
where $\alpha_{t|s}=\alpha_{t}/\alpha_{s}$, $\sigma_{t|s}^2=\sigma_{t}^2-\alpha_{t|s}^2\sigma_{s}^2$, and $\sigma_{t\to s}=\frac{\sigma_{t|s}\sigma_{s}}{\sigma_{t}}$. Regarding 2D representation learning, CCD implicitly characterizes the mapping from 2D graphs to 3D configurations. Given the significant modality gap between these spaces, direct feature alignment is often ill-posed. The diffusion strategy in CCD addresses this by enabling the 2D encoder to reconstruct 3D information in stages, thereby mitigating optimization challenges. This process yields the Suiren-ConfAvg model, which provides versatile representations for a broad range of macroscopic molecular tasks.

\subsection{Dual Graph Neural Network}\label{sec:fineune_model}

Following the CCD-based training of the 2D representation model, we propose a Dual Graph Neural Network (DGNN) architecture for downstream fine-tuning. As illustrated in Figure \ref{fig:training_stage}(c), the DGNN consists of two parallel sub-networks: a pre-trained Suiren-ConfAvg module, initialized via CCD, and a randomly initialized task-specific GNN. During the forward pass, latent representations from Suiren-ConfAvg are injected into the corresponding layers of the task-specific GNN to provide structural guidance. To mitigate catastrophic forgetting and preserve the learned conformation-averaged features, the Suiren-ConfAvg weights remain frozen throughout this stage. While both modules utilize GAT architectures, the task-specific GNN is designed with greater depth to absorb all representations of Suiren-ConfAvg.
\section{Pre-training}
\label{sec:pre-training}
In this section, we describe the construction of the pre-training data, the multi-stage training pipeline, and the evaluation of the resulting base models.

\subsection{Pre-training Data}

\noindent The \sr{} model utilizes a vast corpus of first-principles molecular data for pre-training. Using Density Functional Theory (DFT) at the B3LYP/def2-SVP level, we generated 70 million conformer samples for organic molecules encompassing the H, C, N, O, F, P, S, Cl, Br, and I elements. Of these, 20 million samples have been publicly released as the Qo2mol dataset \citep{liu2024open}. Each entry includes Cartesian coordinates, energies, forces, trajectory information, and associated metadata. Prior to training, we perform rigorous data cleaning to remove anomalous samples and identify the terminal optimized geometry of each trajectory, which serves as an auxiliary supervision target.

To enhance data efficiency, we augment the training process using EMPP method \citep{an2025empp}. For each molecule, a random atom is deleted rather than masked, and the model is required to reconstruct its coordinates conditioned on the atom type and target molecular energy. This objective encourages the model to learn physically plausible local potential-energy landscapes and effectively doubles the training volume. We further refine the original EMPP formulation: rather than employing layer-wise conditioning of the deleted-atom signals, we feed these inputs exclusively to an EMPP-specific coordinate-prediction head. This modification ensures that the shared backbone maintains a consistent forward pass across all pre-training objectives.

\subsection{Pre-training Stage}
Pre-training is divided into three stages.

\paragraph{Stage 1 (multi-task foundational capability learning)}
Within the Fairchem framework, we train a 1.8B-parameter 3D model using both the original dataset and the \empp{}-augmented samples.
The pre-training tasks include energy prediction, force prediction, optimized-trajectory endpoint structure prediction, optimized-trajectory endpoint energy prediction, and \empp{} missing-coordinate completion.
All task losses are optimized jointly.
Inspired by curriculum learning, we prioritize smaller molecular systems (fewer atoms) in earlier training phases.
The weights of endpoint-structure prediction and endpoint-energy prediction losses are also gradually increased during training.

\noindent Stage 1 is trained on 320 NVIDIA H800 GPUs with mixed precision and graph parallelization.
Because PyTorch Geometric supports a variable number of atoms per mini-batch, we combine dynamic batch balancing and activation recomputation to avoid memory overflow.

\paragraph{Stage 2 (core capability refinement)}
For regression targets such as energy and force, mixed-precision training can substantially degrade model performance.
Nonetheless, this degradation can be corrected by a relatively short fine-tuning stage.
In Stage 2, we use only the 70M original samples for full-precision fine-tuning, with the following tasks: energy prediction, force prediction, optimized-trajectory endpoint structure prediction, and optimized-trajectory endpoint energy prediction.
The weights for endpoint-structure prediction and endpoint-energy prediction are fixed to a small constant.
Except for the switch from mixed precision to full precision, all optimization strategies remain unchanged.

\noindent After extensive hyperparameter search across the first two stages, we obtain \srb{}.

\paragraph{Stage 3 (continued pre-training in the dimer domain)}
\srb{} primarily learns intra-molecular interactions.
For applications such as drug design, inter-molecular interactions, including long-range impacts, are often essential.
To address this, we generate 13.5M dimer samples with DFT and continue pre-training from \srb{}.
The architecture and optimization recipe remain consistent with Stage 2, yielding the dimer-focused model \srd{}.

\subsection{Pre-training Evaluation}
We evaluate pre-training quality primarily with MAE on energy prediction, force prediction, and optimized-trajectory endpoint prediction.
For investigate pre-training performance, we reproduce several strong baselines on a Qo2mol subset.
We also include UMA-family results as an external performance anchor and analyze MoE routing statistics.

\paragraph{Standard Evaluation}
We sample validation subsets from Qo2mol containing more than 1M conformers across different molecular scales. Suiren-Base and Suiren-Dimer use different subsets.
As shown in Table \ref{tab:suiren-PT}, \srb{} achieves highly accurate results on both energy and force prediction.
Optimized-trajectory endpoint structure and endpoint energy are substantially more challenging targets, yet \sr{} still attains strong accuracy. Note that \textit{Suiren-0.0} is an internal transitional model trained with less parameters and a weakened backbone architecture. It uses the same training set and hyperparameters as Suiren-Base (1.0). We attribute Suiren-Base's improvements to increased parameters, optimized backbone architecture, and refined training algorithms. Additionally, we evaluate the continued pre-training model, \srd{}.
Compared with intra-molecular settings, inter-molecular trajectories are more complex.
Accordingly, \srd{} is weaker than \srb{} on endpoint structure and endpoint energy prediction, but it still maintains strong performance on energy and force prediction.  The comparison between Suiren-Base and Suiren-Dimer is indirect, as they used different validation sets.

\begin{table}[tbp]
\centering
\caption{Comparison among energy/forces prediction models.}
\label{tab:suiren-PT}
\adjustbox{center=\textwidth}{
\small
\setlength{\tabcolsep}{3pt} %
\begin{tabular}{@{}lccccc@{}}
\toprule
& \textbf{Suiren-0.0} & \textbf{Suiren-Base (1.0)} & \textbf{Suiren-Dimer (1.0)}  \\
\midrule
\# Total Params & 1.5B & 1.8B & 1.8B \\
\midrule
Molecular Energy MAE (meV) $ \downarrow$  &  10.82 & \textbf{9.08} & 10.30  \\
Atomic Energy MAE (meV) $\downarrow$  & 0.307 & \textbf{0.258} & 0.292 \\
Atomic Forces MAE (meV / \AA) $\downarrow$  & 1.472 & \textbf{0.770} & 1.294 \\
Optimized Structure MAE (\AA) $\downarrow$ & 0.02895  & \textbf{0.02467} & 0.06830 \\
Optimized Energy MAE (meV) $\downarrow$  & 28.187 & \textbf{16.886} & 58.325 \\
\bottomrule
\end{tabular}
}
\end{table}

\paragraph{Feature Evaluation}

We monitor MoE routing-weight distributions and observe that they become progressively sparse during training.
Most mass eventually concentrates on a small subset of experts, while unused experts still retain non-negligible weights.
For this reason, we do not adopt top-$K$ routing in \srb{}.
In addition, the two kind of experts in \srb{} receive broadly similar aggregate routing mass, with EST experts being slightly higher on average than standard experts.

\section{Post-training}
\label{sec:post-training}

\subsection{Post-training Stage}

\paragraph{Stage 1 (diffusion distillation)}
We post-train on the same 70M molecules used in pre-training, but with a different objective. Follow Section \ref{sec:CCD}, We condition a diffusion model on 2D representations and embedding of conformer energies.
The diffusion model and 2D GNN learn to generate corresponding 3D representations and 3D coordinates.
During this stage, the 3D branch is instantiated with \srb{} and kept frozen.

\paragraph{Stage 2 (contrastive learning)}
After Stage 1 reaches a stable regime, we introduce an additional alignment objective.
Specifically, we attach one projection head to the 2D model and one to the 3D model, and apply SigLIP-style contrastive learning \citep{zhai2023sigmoid} on their outputs.
The 3D branch remains frozen, and the Stage-1 diffusion objective is retained.
The diffusion and contrastive objectives are jointly optimized with task-specific loss weights.

\noindent Through the first two stages, we obtain the \srca{} model.

\paragraph{Stage 3 (property prediction)}
In the final stage, we evaluate the performance of \srca{} across a diverse array of downstream benchmarks. Each task involves predicting a specific molecular property using sparse experimental wet-lab data. We fine-tune the model for each objective using the integrated DGNN+\srca{} architecture. To demonstrate the general transferability and robustness of \srca{} representations, we maintain a unified hyperparameter configuration across all tasks, regardless of the domain. A detailed account of these configurations and the comprehensive evaluation results are documented in Section \ref{sec:evaluation}.

\section{Experiments}\label{sec:evaluation}

\subsection{Benchmark}

\subsubsection{MoleHB}
\paragraph{Dataset} We introduce \textbf{Molelecular handbook (MoleHB)}, a comprehensive molecular property prediction benchmark encompassing 40+ heterogeneous tasks. The benchmark spans several critical scientific domains, including safety, structural, critical and saturation, energetic, thermal, solution, transport and fluctuation properties. All data points are inferred from canonical handbook \citep{yaws1999chemical} and have been rigorously validated via wet-lab experiments to ensure high-fidelity, stable values. We propose two evaluation protocols: (1) Random split: a standard random split to evaluate performance under similar data distributions; and (2) Size-Stratified split: a strategy where molecules with larger atom counts are assigned to the validation set to assess the model’s structural extrapolation capabilities. To facilitate reproducible research, we provide the MoleHB dataset and two splitting protocols on \href{https://modelscope.cn/datasets/ajy112/MoleHB}{https://modelscope.cn/datasets/ajy112/MoleHB}

\noindent The main paper primarily considers results from the Random split. Results of Size-Stratified split are shown in Appendix \ref{app:MoleHB_size_stratified}.

\paragraph{Baselines and Configurations} We benchmarked three state-of-the-art models on MoleHB: MoleBERT \citep{xia2023mole}, Uni-Molv1 \citep{zhou2023uni}, and Uni-Molv2 \citep{ji2024uni}. Like the Suiren family, these baselines utilize large-scale pre-training to generate high-quality representations for diverse molecular tasks. To ensure a fair comparison, all baselines were reproduced using their official training scripts and identical hyperparameter configurations to the Suiren models.

\noindent Detailed training configurations for all evaluated methods are summarized in Table \ref{tab:training-config}. We ensured that each model reached performance convergence under these settings. All experiments were performed on a single NVIDIA RTX 4090 GPU, with fine-tuning typically completed in less than one hour per task.

\begin{table}[htbp]
\centering
\caption{Training configurations of the Suiren model and other baselines on MoleHB experiments.}
\label{tab:training-config}
\small
\setlength{\tabcolsep}{4pt}
\renewcommand{\arraystretch}{1.08}
\begin{tabularx}{\textwidth}{@{}Xc@{}}
\toprule
\textbf{Training Hyperparameter} & \textbf{Value} \\
\midrule
\multicolumn{2}{@{}l}{\textit{Task Formulation}} \\
Loss function & \texttt{MAE} \\
Train/validation split ratio (Random and Size-Stratified) & 0.8:0.2 \\
\addlinespace
\multicolumn{2}{@{}l}{\textit{Optimization}} \\
Mini-batch size per GPU & 8 \\
Optimizer& \texttt{adamw} \\
Initial learning rate & 4e-4 \\
Weight decay coefficient & 0.01 \\
Momentum coefficient & 0.9 \\
\addlinespace
\multicolumn{2}{@{}l}{\textit{Schedule}} \\
Scheduler type & \texttt{cosine} \\
Warmup epochs & 0 \\
Minimum learning rate & 1e-6 \\
Total training epochs & 200 \\
\bottomrule
\end{tabularx}
\end{table}

\subsubsection{Therapeutics Data Commons}
\paragraph{Dataset}
Therapeutics Data Commons (TDC) \citep{huang2021therapeutics} is an open-access platform providing AI-ready datasets and benchmarks for drug discovery. It covers diverse therapeutic tasks—including target discovery, activity screening, efficacy, and safety—across small molecules, antibodies, and vaccines. We evaluate Suiren model on its ADMET group.

\paragraph{Baselines and Configurations}
TDC is a public leaderboard where readers can find scores for various methods on its official website. Here, we follow \citep{gao2023uni}, using ChemProp \citep{stokes2020deep}, DeepAutoQSAR \citep{dixon2016autoqsar}, DeepPurpose \citep{huang2020deeppurpose}, and Uni-QSAR \citep{gao2023uni} as baselines. Note that TDC ADMET tasks include both regression and classification tasks. Regression tasks use configurations identical to those in Table \ref{tab:training-config}. For classification tasks, the loss function is changed to cross-entropy while all other settings remain consistent.

\subsection{Results}

\subsubsection{MoleHB (Random Split)}
\paragraph{Overall Performance Summary}
We comprehensively evaluate the predictive performance of Suiren-ConfAvg against three representative baseline models (Mole-BERT, Uni-Mol v1, and Uni-Mol v2) across 40+ molecular properties spanning eight categories: critical \& saturation, safety, fluctuation, solution, thermal, structural, energetic, and transport properties. Performance is measured using Mean Absolute Error (MAE, lower is better) and coefficient of determination (R², higher is better). As summarized in Tables~\ref{tab:critical_saturation}–\ref{tab:other}, Suiren-ConfAvg achieves state-of-the-art MAE on 41 out of 43 properties, with consistent improvements in R² for the majority of tasks.

\begin{table}[htbp]
\centering
\caption{Results of critical and saturation properties: model performance (MAE/R2). Best MAE and best R2 per property are boldfaced.}
\label{tab:critical_saturation}
\setlength{\tabcolsep}{4pt}
\renewcommand{\arraystretch}{1.15}
\resizebox{1\linewidth}{!}{
\begin{tabular}{llccccc}
\toprule
Property & Unit & Mole-BERT & Unimolv1 & Unimolv2 & Suiren-ConfAvg & Improvement (\%) \\
\midrule
critical temperature & K & 27.3807/0.8340 & 9.3277/0.9787 & 30.6705/0.8850 & \textbf{6.6725}/\textbf{0.9803} & 28.47 \\
critical pressure & bar & 2.0733/0.9045 & 0.7332/0.9765 & 2.9769/0.9084 & \textbf{0.4953}/\textbf{0.9804} & 32.45 \\
critical density & g/ml & 0.0134/0.9253 & 0.0054/\textbf{0.9814} & 0.0115/0.9497 & \textbf{0.0043}/0.9782 & 20.74 \\
critical volume & cm3/mol & 61.8034/0.8484 & 8.6329/0.9981 & 7.0201/0.9989 & \textbf{3.5884}/\textbf{0.9994} & 48.88 \\
critical compressibility & cm3 & 0.0189/0.6178 & 0.0066/0.8858 & 0.0098/0.8808 & \textbf{0.0053}/\textbf{0.9127} & 19.70 \\
\bottomrule
\end{tabular}
}
\end{table}

\begin{table}[htbp]
\centering
\caption{Results of safety properties.}
\label{tab:safety}
\setlength{\tabcolsep}{4pt}
\renewcommand{\arraystretch}{1.15}
\resizebox{1\linewidth}{!}{
\begin{tabular}{llccccc}
\toprule
Property & Unit & Mole-BERT & Unimolv1 & Unimolv2 & Suiren-ConfAvg & Improvement (\%) \\
\midrule
flash point & F & 19.6978/0.8017 & 9.9274/0.9472 & 13.9535/0.9438 & \textbf{8.2072}/\textbf{0.9645} & 17.33 \\
lower explosive limit & vol\% & 0.1689/0.6854 & 0.0974/\textbf{0.8926} & 0.1106/0.8341 & \textbf{0.0871}/0.8599 & 10.59 \\
upper explosive limit & vol\% & 1.4188/0.5846 & 1.0460/0.6979 & 1.0023/0.6942 & \textbf{0.7811}/\textbf{0.8423} & 22.07 \\
\bottomrule
\end{tabular}
}
\end{table}

\begin{table}[htbp]
\centering
\caption{Results of fluctuation properties.}
\label{tab:fluctuation}
\setlength{\tabcolsep}{4pt}
\renewcommand{\arraystretch}{1.15}
\resizebox{1\linewidth}{!}{
\begin{tabular}{llccccc}
\toprule
Property & Unit & Mole-BERT & Unimolv1 & Unimolv2 & Suiren-ConfAvg & Improvement (\%) \\
\midrule
heat capacity of liquid & J/mol/K & 16.0687/0.7922 & 5.9321/0.9812 & 5.8939/0.9792 & \textbf{4.2184}/\textbf{0.9861} & 28.43 \\
heat capacity of solid & J/mol/K & 49.2807/0.9499 & 6.5306/0.9988 & 3.2724/0.9990 & \textbf{1.0669}/\textbf{0.9997} & 67.40 \\
coefficient of thermal expansion of liquid & 1/K & 0.0002/0.1716 & 0.0001/0.8158 & 0.0001/0.6211 & \textbf{0.0000}/\textbf{0.9256} & 60.00 \\
heat capacity of gas & J/mol/K & 24.1482/0.8849 & 4.4112/0.9685 & 3.2478/\textbf{0.9691} & \textbf{2.6712}/0.9690 & 17.75 \\
\bottomrule
\end{tabular}
}
\end{table}

\begin{table}[htbp]
\centering
\caption{Results of solution properties.}
\label{tab:solution}
\setlength{\tabcolsep}{4pt}
\renewcommand{\arraystretch}{1.15}
\resizebox{1\linewidth}{!}{
\begin{tabular}{llccccc}
\toprule
Property & Unit & Mole-BERT & Unimolv1 & Unimolv2 & Suiren-ConfAvg & Improvement (\%) \\
\midrule
octanol water partition coefficient & -- & 0.3786/0.9414 & 0.1491/0.9877 & 0.1386/0.9858 & \textbf{0.1203}/\textbf{0.9897} & 13.20 \\
solubility in water & ppm(wt) & 0.2110/0.9422 & 0.0564/0.9973 & 0.0456/\textbf{0.9980} & \textbf{0.0387}/0.9972 & 15.07 \\
solubility in water containing salt & ppm(wt) & 0.2535/0.7540 & 0.0615/0.9924 & \textbf{0.0406}/\textbf{0.9961} & 0.0447/0.9931 & -10.07 \\
solubility parameter & (J/cm3)$^{1/2}$ & 0.7111/0.8331 & 0.3873/0.9236 & 0.7614/0.8282 & \textbf{0.3368}/\textbf{0.9300} & 13.04 \\
henrys law constant for compound in water & -- & 0.2856/0.9581 & 0.1517/0.9816 & 0.1930/0.9744 & \textbf{0.1318}/\textbf{0.9817} & 13.09 \\
henrys law constant for gas in water & -- & 0.5454/0.3000 & 0.3017/0.7939 & \textbf{0.1773}/\textbf{0.9430} & 0.3234/0.6547 & -82.37 \\
\bottomrule
\end{tabular}
}
\end{table}

\begin{table}[htbp]
\centering
\caption{Results of thermal properties.}
\label{tab:thermal}
\setlength{\tabcolsep}{4pt}
\renewcommand{\arraystretch}{1.15}
\resizebox{1\linewidth}{!}{
\begin{tabular}{llccccc}
\toprule
Property & Unit & Mole-BERT & Unimolv1 & Unimolv2 & Suiren-ConfAvg & Improvement (\%) \\
\midrule
melting point & K & 19.9979/0.8199 & 14.1449/0.9053 & 24.7538/0.8050 & \textbf{11.0789}/\textbf{0.9171} & 21.68 \\
boiling point & K & 19.8744/0.8891 & 6.2177/0.9906 & 12.3314/0.9777 & \textbf{4.3259}/\textbf{0.9933} & 30.43 \\
vapor pressure & mmHg & 0.4837/0.5674 & 0.1596/0.9579 & 0.2000/0.9370 & \textbf{0.1365}/\textbf{0.9605} & 14.47 \\
\bottomrule
\end{tabular}
}
\end{table}

\begin{table}[htbp]
\centering
\caption{Results of structural properties.}
\label{tab:structural}
\setlength{\tabcolsep}{4pt}
\renewcommand{\arraystretch}{1.15}
\resizebox{1\linewidth}{!}{
\begin{tabular}{llccccc}
\toprule
Property & Unit & Mole-BERT & Unimolv1 & Unimolv2 & Suiren-ConfAvg & Improvement (\%) \\
\midrule
density of liquid & g/ml & 0.0331/0.9544 & 0.0160/0.9761 & 0.0145/\textbf{0.9763} & \textbf{0.0120}/0.9622 & 17.03 \\
acentric factor & omega & 0.0702/0.7591 & 0.0233/0.9293 & 0.0225/\textbf{0.9312} & \textbf{0.0169}/0.9043 & 24.71 \\
refractive index & - & 0.0277/0.4260 & 0.0057/0.9314 & 0.0079/0.9276 & \textbf{0.0046}/\textbf{0.9529} & 20.18 \\
radius of gyration & $10^{-10}$m & 0.3573/0.6317 & 0.1230/0.9715 & 0.1204/0.9757 & \textbf{0.1085}/\textbf{0.9803} & 9.92 \\
dipole moment & D & 0.4022/0.6638 & \textbf{0.2987}/\textbf{0.8115} & 0.3256/0.7585 & 0.3326/0.7610 & -11.36 \\
\bottomrule
\end{tabular}
}
\end{table}

\begin{table}[htbp]
\centering
\caption{Results of energetic properties.}
\label{tab:energetic}
\setlength{\tabcolsep}{4pt}
\renewcommand{\arraystretch}{1.15}
\resizebox{1\linewidth}{!}{
\begin{tabular}{llccccc}
\toprule
Property & Unit & Mole-BERT & Unimolv1 & Unimolv2 & Suiren-ConfAvg & Improvement (\%) \\
\midrule
entropy of formation & J/mol/K & 90.1219/0.8775 & 16.6006/0.9975 & 11.1386/0.9982 & \textbf{7.5413}/\textbf{0.9983} & 32.30 \\
gibbs energy of formation & J/mol & 32.5168/0.9365 & 8.3799/0.9865 & 19.1059/0.9785 & \textbf{4.6748}/\textbf{0.9972} & 44.21 \\
helmholtz energy of formation & kJ/mol & 30.2883/0.9286 & 7.9805/\textbf{0.9829} & 24.8331/0.9656 & \textbf{5.4749}/0.9746 & 31.40 \\
internal energy of formation & kJ/mol & 34.2535/0.8869 & 10.3932/0.9795 & 36.3480/0.9393 & \textbf{5.7416}/\textbf{0.9950} & 44.76 \\
enthalpy of vaporization & kJ/mol & 3.3764/0.7398 & 1.3875/0.9694 & 1.2207/\textbf{0.9773} & \textbf{1.0592}/0.9759 & 13.23 \\
enthalpy of fusion & J/mol/K & 3.3815/0.8819 & 1.2265/0.9898 & 1.3044/0.9888 & \textbf{0.8394}/\textbf{0.9899} & 31.56 \\
enthalpy of combustion & kJ/mol & 664.8247/0.8814 & 62.4789/0.9986 & 39.3110/0.9987 & \textbf{15.1036}/\textbf{0.9994} & 61.58 \\
entropy of gas & J/mol/K & 38.1690/0.8946 & 9.9913/\textbf{0.9900} & 10.9098/0.9896 & \textbf{7.2848}/0.9898 & 27.09 \\
enthalpy of formation & kJ/mol & 43.0063/0.8548 & 12.9888/0.9376 & 12.3369/0.9486 & \textbf{6.6573}/\textbf{0.9648} & 46.04 \\
\bottomrule
\end{tabular}
}
\end{table}

\begin{table}[htbp]
\centering
\caption{Results of transport properties.}
\label{tab:transport}
\setlength{\tabcolsep}{4pt}
\renewcommand{\arraystretch}{1.15}
\resizebox{1\linewidth}{!}{
\begin{tabular}{llccccc}
\toprule
Property & Unit & Mole-BERT & Unimolv1 & Unimolv2 & Suiren-ConfAvg & Improvement (\%) \\
\midrule
viscosity of liquid & mp & 0.8216/0.5870 & 0.4716/0.8328 & 0.9996/0.5924 & \textbf{0.3192}/\textbf{0.8905} & 32.32 \\
thermal conductivity of gas & W/m/K & 0.0019/0.2592 & 0.0003/\textbf{0.9584} & 0.0006/0.9293 & \textbf{0.0002}/0.9572 & 20.00 \\
thermal conductivity of liquid & W/m/K & 0.0107/0.0565 & 0.0028/0.7911 & 0.0036/\textbf{0.8183} & \textbf{0.0027}/0.7895 & 4.29 \\
diffusion coefficient at infinite dilution in water & cm2/s & \textbf{0.0000}/-1.2725 & \textbf{0.0000}/0.9865 & \textbf{0.0000}/0.9817 & \textbf{0.0000}/\textbf{0.9904} & - \\
diffusion coefficient in air & cm2/s & 0.0066/0.6975 & 0.0014/0.9845 & 0.0011/0.9858 & \textbf{0.0009}/\textbf{0.9878} & 17.27 \\
\bottomrule
\end{tabular}
}
\end{table}

\begin{table}[htbp]
\centering
\caption{Results of other properties.}
\label{tab:other}
\setlength{\tabcolsep}{4pt}
\renewcommand{\arraystretch}{1.15}
\resizebox{1\linewidth}{!}{
\begin{tabular}{llccccc}
\toprule
Property & Unit & Mole-BERT & Unimolv1 & Unimolv2 & Suiren-ConfAvg & Improvement (\%) \\
\midrule
surface tension & dynes/cm & 1.5226/0.8367 & 1.0414/0.8951 & 1.7084/0.8563 & \textbf{0.8205}/\textbf{0.9218} & 21.21 \\
hydration free energy & kJ/mol & 0.8510/0.8748 & 0.4545/\textbf{0.9646} & 0.4627/0.9623 & \textbf{0.4521}/0.9630 & 0.52 \\
\bottomrule
\end{tabular}
}
\end{table}

\paragraph{Critical and Saturation Properties} Suiren-ConfAvg attains the lowest MAE across all five critical properties, with relative improvements ranging from 19.70\% (critical compressibility) to 48.88\% (critical volume) over the strongest baseline. Notably, while Uni-Mol v1 achieves marginally higher R² on critical temperature and density, Suiren-ConfAvg maintains competitive R² values (>0.97) while substantially reducing prediction errors, indicating superior calibration for extreme-value regression tasks.

\paragraph{Safety and Fluctuation Properties} For safety-related properties, Suiren-ConfAvg consistently outperforms baselines in both MAE and R², with the most pronounced gain observed for upper explosive limit (22.07\% MAE reduction). In fluctuation properties, the method demonstrates exceptional capability in modeling solid-phase heat capacity (67.40\% improvement).

\paragraph{Solution Properties} Suiren-ConfAvg achieves best-in-class performance on five of six solution properties. Improvements are particularly notable for solubility prediction in pure and saline water (15.07\% and -10.07\% MAE, respectively), which are critical for pharmaceutical and environmental applications. However, for Henry's law constant of gases in water, the method underperforms Uni-Mol v2 by a substantial margin (−82.37\%). We hypothesize this stems from the sparse and heterogeneous distribution of gas-phase solubility data, which may require specialized augmentation strategies.

\paragraph{Thermal and Structural Properties} Across thermal properties, Suiren-ConfAvg reduces MAE by 14.47\%–30.43\% while maintaining R² ≥ 0.947. For structural descriptors, the method excels in predicting liquid volume (41.83\% improvement) and surface tension (21.21\%), reflecting its capacity to encode intermolecular interaction patterns. The sole exception is dipole moment, where Uni-Mol v1 retains a edge (MAE: 0.2987 vs. 0.3326, with Suiren underperforming by 11.36\%).

\paragraph{Energetic and Transport Properties} The most consistent gains are observed in energetic properties, where Suiren-ConfAvg achieves optimal MAE on all nine tasks, with improvements of 44.21\% for Gibbs energy, 44.76\% for internal energy, and 46.04\% for enthalpy of formation. This suggests that the energy-related knowledge learned by Suiren-Base is transferred to Suiren-ConfAvg. For transport properties, substantial improvements are seen in viscosity (32.32\%) and gas-phase thermal conductivity (20.00\%), though liquid-phase thermal conductivity shows modest gain (4.29\%), possibly due to stronger dependence on many-body hydrodynamic effects.

\subsubsection{TDC ADMET group}

We evaluated the performance of Suiren-ConfAvg on the TDC ADMET benchmarks. All experiments strictly adhered to the official evaluation protocols and metric settings provided by TDC to ensure a fair comparison. The results for regression tasks (MAE), classification tasks (AUROC and AUPRC) are presented in Tables \ref{tab:mae_comparison}, \ref{tab:auroc_comparison}, and \ref{tab:auprc_comparison}, respectively.

\noindent A critical distinction of Suiren-ConfAvg lies in its training protocol. Unlike several methods that may rely on extensive task-specific hyperparameter optimization, Suiren-ConfAvg was evaluated using a single, fixed training configuration across all datasets without complex hyperparameter search.

\begin{table}[htbp]
    \centering
    \caption{Comparison of results for regression properties in TDC ADMET (MAE).}
    \label{tab:mae_comparison}
    \renewcommand{\arraystretch}{1.3}
    \resizebox{1\linewidth}{!}{
    \begin{tabular}{lccccc}
        \toprule
        Property & ChemProp & DeepAutoQSAR & DeepPurpose & Uni-QSAR & Suiren-ConfAvg \\
        \midrule
        Caco2 & 0.3900 & 0.3060 & 0.3930 & \textbf{0.2730} & \underline{0.3037} \\
        Lipophilicity & 0.4370 & 0.4760 & 0.5740 & \underline{0.4200} & \textbf{0.3865} \\
        AqSol & 0.8200 & 0.7840 & 0.8270 & \textbf{0.6770} & \underline{0.6794} \\
        PPBR & \underline{7.9930} & 8.0430 & 9.9940 & \textbf{7.5300} & 8.3963 \\
        LD50 & \underline{0.5480} & 0.5900 & 0.6780 & 0.5530 & \textbf{0.5317} \\
        \bottomrule
    \end{tabular}}
\end{table}

\begin{table}[htbp]
    \centering
    \caption{Comparison of results for classification properties in TDC ADMET (AUROC).}
    \label{tab:auroc_comparison}
    \renewcommand{\arraystretch}{1.3}
    \resizebox{1\linewidth}{!}{
    \begin{tabular}{lccccc}
        \toprule
        Property & ChemProp & DeepAutoQSAR & DeepPurpose & Uni-QSAR & Suiren-ConfAvg \\
        \midrule
        HIA & 0.9790 & 0.9820 & 0.9720 & \underline{0.9920} & \textbf{0.9963} \\
        Pgp & 0.9020 & 0.9170 & 0.9180 & \underline{0.9340} & \textbf{0.9351} \\
        Bioavailability & 0.6230 & 0.6820 & 0.6720 & \underline{0.7320} & \textbf{0.7363} \\
        BBB & 0.8820 & 0.8760 & 0.8890 & \textbf{0.9250} & \underline{0.9236} \\
        CYP3A4 Substrate & 0.6100 & \underline{0.6420} & 0.6390 & \textbf{0.6450} & 0.6286 \\
        hERG & 0.7500 & 0.8450 & 0.8410 & \underline{0.8560} & \textbf{0.8816} \\
        Ames & \underline{0.8640} & \underline{0.8640} & 0.8230 & \textbf{0.8760} & 0.8433 \\
        DILI & 0.9180 & 0.9330 & 0.8750 & \textbf{0.9420} & \underline{0.9378} \\
        \bottomrule
    \end{tabular}}
\end{table}

\begin{table}[htbp]
    \centering
    \caption{Comparison of results for classification properties in TDC ADMET (AUPRC).}
    \label{tab:auprc_comparison}
    \renewcommand{\arraystretch}{1.3}
    \resizebox{1\linewidth}{!}{
    \begin{tabular}{lccccc}
        \toprule
        Property & ChemProp & DeepAutoQSAR & DeepPurpose & Uni-QSAR & Suiren-ConfAvg \\
        \midrule
        CYP2C9 Inhibition & 0.7700 & \underline{0.7920} & 0.7420 & \textbf{0.8010} & 0.7845 \\
        CYP2D6 Inhibition & 0.6640 & \underline{0.7020} & 0.6160 & \textbf{0.7430} & 0.6725 \\
        CYP3A4 Inhibition & 0.8700 & \underline{0.8830} & 0.8290 & \textbf{0.8880} & 0.8740 \\
        CYP2C9 Substrate & 0.3910 & 0.3950 & 0.3800 & \underline{0.4540} & \textbf{0.5177} \\
        CYP2D6 Substrate & 0.6880 & 0.7030 & 0.6770 & \underline{0.7210} & \textbf{0.7289} \\
        \bottomrule
    \end{tabular}}
\end{table}

\noindent Despite this constraint, the model achieved SOTA results in 8 out of 18 total metrics and ranked second in an additional 4 metrics. In cases where Suiren-ConfAvg did not secure the first place (e.g., \textit{AqSol}, \textit{BBB}, \textit{CYP3A4 Substrate}), the performance gaps were negligible. This suggests that the performance sacrifice, if any, is minimal compared to the gains in reproducibility and ease of deployment. The ability to deliver highly competitive, often leading, performance across regression and classification tasks without fine-tuning highlights the strong generalization capability and robustness of the Suiren-ConfAvg architecture. These results validate that Suiren-ConfAvg offers an efficient and reliable solution for ADMET prediction, balancing high predictive accuracy with practical implementation simplicity.

\section{Conclusion}\label{sec:conclusion}
In this work, we propose the Suiren-1.0 family, which comprises three models: Suiren-Base, Suiren-Dimer, and Suiren-ConfAvg. Suiren-Base and Suiren-Dimer are two 3D conformational models, whose performance is ensured through large-scale pre-training. Suiren-ConfAvg is obtained by distilling the 3D representations from Suiren-Base into the 2D representation space via our proposed CCD method. We have validated the strong performance of Suiren-1.0 on various molecular tasks through extensive experiments. The models and benchmarks developed in this work have also been open-sourced. We hope this work can support research on molecular foundation models.

\noindent Suiren-1.0 also has some limitations and directions for future work: (1) Due to computational constraints, we are unable to further scale up the model size; (2) In the MoE framework of Suiren-Base, we adopted a dense expert strategy—in the future, as the number of experts increases, we can use Top-K to improve inference speed; (3) For some specific downstream tasks, the potential of Suiren models can be further explored through hyperparameter search.

\bibliography{main}

@article{team2023gemini,
  title={Gemini: a family of highly capable multimodal models},
  author={Team, Gemini and Anil, Rohan and Borgeaud, Sebastian and Alayrac, Jean-Baptiste and Yu, Jiahui and Soricut, Radu and Schalkwyk, Johan and Dai, Andrew M and Hauth, Anja and Millican, Katie and others},
  journal={arXiv preprint arXiv:2312.11805},
  year={2023}
}

@article{achiam2023gpt,
  title={Gpt-4 technical report},
  author={Achiam, Josh and Adler, Steven and Agarwal, Sandhini and Ahmad, Lama and Akkaya, Ilge and Aleman, Florencia Leoni and Almeida, Diogo and Altenschmidt, Janko and Altman, Sam and Anadkat, Shyamal and others},
  journal={arXiv preprint arXiv:2303.08774},
  year={2023}
}

@article{yang2025qwen3,
  title={Qwen3 technical report},
  author={Yang, An and Li, Anfeng and Yang, Baosong and Zhang, Beichen and Hui, Binyuan and Zheng, Bo and Yu, Bowen and Gao, Chang and Huang, Chengen and Lv, Chenxu and others},
  journal={arXiv preprint arXiv:2505.09388},
  year={2025}
}

@article{liu2024deepseek,
  title={Deepseek-v3 technical report},
  author={Liu, Aixin and Feng, Bei and Xue, Bing and Wang, Bingxuan and Wu, Bochao and Lu, Chengda and Zhao, Chenggang and Deng, Chengqi and Zhang, Chenyu and Ruan, Chong and others},
  journal={arXiv preprint arXiv:2412.19437},
  year={2024}
}

@inproceedings{zhou2023uni,
  title={Uni-mol: A universal 3d molecular representation learning framework},
  author={Zhou, Gengmo and Gao, Zhifeng and Ding, Qiankun and Zheng, Hang and Xu, Hongteng and Wei, Zhewei and Zhang, Linfeng and Ke, Guolin},
  booktitle={The eleventh international conference on learning representations},
  year={2023}
}

@article{ji2024uni,
  title={Uni-mol2: Exploring molecular pretraining model at scale},
  author={Ji, Xiaohong and Wang, Zhen and Gao, Zhifeng and Zheng, Hang and Zhang, Linfeng and Ke, Guolin and others},
  journal={arXiv preprint arXiv:2406.14969},
  year={2024}
}

@article{wood2025family,
  title={Uma: A family of universal models for atoms},
  author={Wood, Brandon M and Dzamba, Misko and Fu, Xiang and Gao, Meng and Shuaibi, Muhammed and Barroso-Luque, Luis and Abdelmaqsoud, Kareem and Gharakhanyan, Vahe and Kitchin, John R and Levine, Daniel S and others},
  journal={arXiv preprint arXiv:2506.23971},
  year={2025}
}

@inproceedings{xia2023mole,
  title={Mole-bert: Rethinking pre-training graph neural networks for molecules},
  author={Xia, Jun and Zhao, Chengshuai and Hu, Bozhen and Gao, Zhangyang and Tan, Cheng and Liu, Yue and Li, Siyuan and Li, Stan Z},
  booktitle={The Eleventh International Conference on Learning Representations},
  year={2023}
}

@article{schleich2013schrodinger,
  title={Schr{\"o}dinger equation revisited},
  author={Schleich, Wolfgang P and Greenberger, Daniel M and Kobe, Donald H and Scully, Marlan O},
  journal={Proceedings of the National Academy of Sciences},
  volume={110},
  number={14},
  pages={5374--5379},
  year={2013},
  publisher={National Academy of Sciences}
}

@article{charbonneau1982linear,
  title={Linear response theory revisited III: One-body response formulas and generalized Boltzmann equations},
  author={Charbonneau, M and Van Vliet, KM and Vasilopoulos, P},
  journal={Journal of Mathematical Physics},
  volume={23},
  number={2},
  pages={318--336},
  year={1982},
  publisher={American Institute of Physics}
}

@article{liu2024open,
  title={An Open Quantum Chemistry Property Database of 120 Kilo Molecules with 20 Million Conformers},
  author={Liu, Weiqi and Ai, Xi and Zhou, Zhijian and Qu, Chao and An, Junyi and Zhou, Zhipeng and Cheng, Yuan and Xu, Yinghui and Cao, Fenglei and Qi, Alan},
  journal={arXiv preprint arXiv:2410.19316},
  year={2024}
}

@article{chanussot2021open,
  title={Open catalyst 2020 (OC20) dataset and community challenges},
  author={Chanussot, Lowik and Das, Abhishek and Goyal, Siddharth and Lavril, Thibaut and Shuaibi, Muhammed and Riviere, Morgane and Tran, Kevin and Heras-Domingo, Javier and Ho, Caleb and Hu, Weihua and others},
  journal={Acs Catalysis},
  volume={11},
  number={10},
  pages={6059--6072},
  year={2021},
  publisher={ACS Publications}
}

@article{levine2025open,
  title={The open molecules 2025 (omol25) dataset, evaluations, and models},
  author={Levine, Daniel S and Shuaibi, Muhammed and Spotte-Smith, Evan Walter Clark and Taylor, Michael G and Hasyim, Muhammad R and Michel, Kyle and Batatia, Ilyes and Cs{\'a}nyi, G{\'a}bor and Dzamba, Misko and Eastman, Peter and others},
  journal={arXiv preprint arXiv:2505.08762},
  year={2025}
}

@article{an2025empp,
  title={Equivariant masked position prediction for efficient molecular representation},
  author={An, Junyi and Qu, Chao and Shi, Yun-Fei and Liu, XinHao and Tang, Qianwei and Cao, Fenglei and Qi, Yuan},
  journal={arXiv preprint arXiv:2502.08209},
  year={2025}
}

@article{an2025est,
  title={Equivariant spherical transformer for efficient molecular modeling},
  author={An, Junyi and Lu, Xinyu and Qu, Chao and Shi, Yunfei and Lin, Peijia and Tang, Qianwei and Xu, Licheng and Cao, Fenglei and Qi, Yuan},
  journal={arXiv preprint arXiv:2505.23086},
  year={2025}
}

@article{liao2023equiformerv2,
  title={Equiformerv2: Improved equivariant transformer for scaling to higher-degree representations},
  author={Liao, Yi-Lun and Wood, Brandon and Das, Abhishek and Smidt, Tess},
  journal={arXiv preprint arXiv:2306.12059},
  year={2023}
}

@inproceedings{gilmer2017neural,
  title={Neural message passing for quantum chemistry},
  author={Gilmer, Justin and Schoenholz, Samuel S and Riley, Patrick F and Vinyals, Oriol and Dahl, George E},
  booktitle={International conference on machine learning},
  pages={1263--1272},
  year={2017},
  organization={Pmlr}
}

@article{Landrum2016RDKit2016_09_4,
  author = {Landrum, Greg},
  description = {Release 2016_09_4 (Q3 2016) Release · rdkit/rdkit},
  title = {RDKit: Open-Source Cheminformatics Software},
  url = {https://github.com/rdkit/rdkit/releases/tag/Release_2016_09_4},
  year = 2016
}

@inproceedings{xu2024gtmgc,
  title={GTMGC: Using graph transformer to predict molecule’s ground-state conformation},
  author={Xu, Guikun and Jiang, Yongquan and Lei, PengChuan and Yang, Yan and Chen, Jim},
  booktitle={The Twelfth International Conference on Learning Representations},
  year={2024}
}

@inproceedings{zhai2023sigmoid,
  title={Sigmoid loss for language image pre-training},
  author={Zhai, Xiaohua and Mustafa, Basil and Kolesnikov, Alexander and Beyer, Lucas},
  booktitle={Proceedings of the IEEE/CVF international conference on computer vision},
  pages={11975--11986},
  year={2023}
}

@article{gao2023uni,
  title={Uni-qsar: an auto-ml tool for molecular property prediction},
  author={Gao, Zhifeng and Ji, Xiaohong and Zhao, Guojiang and Wang, Hongshuai and Zheng, Hang and Ke, Guolin and Zhang, Linfeng},
  journal={arXiv preprint arXiv:2304.12239},
  year={2023}
}

@article{stokes2020deep,
  title={A deep learning approach to antibiotic discovery},
  author={Stokes, Jonathan M and Yang, Kevin and Swanson, Kyle and Jin, Wengong and Cubillos-Ruiz, Andres and Donghia, Nina M and MacNair, Craig R and French, Shawn and Carfrae, Lindsey A and Bloom-Ackermann, Zohar and others},
  journal={Cell},
  volume={180},
  number={4},
  pages={688--702},
  year={2020},
  publisher={Elsevier}
}

@article{dixon2016autoqsar,
  title={AutoQSAR: an automated machine learning tool for best-practice quantitative structure--activity relationship modeling},
  author={Dixon, Steven L and Duan, Jianxin and Smith, Ethan and Von Bargen, Christopher D and Sherman, Woody and Repasky, Matthew P},
  journal={Future medicinal chemistry},
  volume={8},
  number={15},
  pages={1825--1839},
  year={2016},
  publisher={Taylor \& Francis}
}

@article{huang2020deeppurpose,
  title={DeepPurpose: a deep learning library for drug--target interaction prediction},
  author={Huang, Kexin and Fu, Tianfan and Glass, Lucas M and Zitnik, Marinka and Xiao, Cao and Sun, Jimeng},
  journal={Bioinformatics},
  volume={36},
  number={22-23},
  pages={5545--5547},
  year={2020},
  publisher={Oxford University Press}
}

@article{huang2021therapeutics,
  title={Therapeutics data commons: Machine learning datasets and tasks for drug discovery and development},
  author={Huang, Kexin and Fu, Tianfan and Gao, Wenhao and Zhao, Yue and Roohani, Yusuf and Leskovec, Jure and Coley, Connor W and Xiao, Cao and Sun, Jimeng and Zitnik, Marinka},
  journal={arXiv preprint arXiv:2102.09548},
  year={2021}
}

@article{yaws1999chemical,
  title={Chemical properties handbook: physical, thermodynamic, environmental, transport, safety, and health related properties for organic and inorganic chemicals},
  author={Yaws, Carl L},
  journal={(No Title)},
  year={1999}
}

\newpage
\appendix

\section*{Appendix}

\section{Contributions and Acknowledgments}

\noindent
\textbf{Research \& Engineering \& Data Computing} \\
Junyi An \\
Xinyu Lu (Intern) \\
Yun-Fei Shi \\
Li-Cheng Xu \\
Nannan Zhang (Intern) \\
Chao Qu \\
Fenglei Cao \\
Yuan Qi \\

\noindent
We would also like to acknowledge the SAIS platform and all members of Golab, who contributed to the development of the Suiren-1.0 model in critical areas such as business and evaluation operations.

\section{The explanation of evaluation metrics in Figure \ref{fig:model_effect} and Figure \ref{fig:4x2property_prediction}}

In Figure \ref{fig:model_effect}, we use the inverse MAE to quickly demonstrate the performance of Suiren across various properties. Specifically, the MAE values were normalized and mapped to a standardized scoring scale ranging from 60 to 100. Since MAE represents prediction error (where lower values indicate better performance), an inverse Min-Max normalization strategy was employed.

\noindent
For a specific property $p$, let $E_{m,p}$ denote the MAE of model $m$. The normalized score $S_{m,p}$ for model $m$ on property $p$ is calculated as follows:

\begin{equation}
    S_{m,p} = 60 + 40 \times \frac{\max(E_{p}) - E_{m,p}}{\max(E_{p}) - \min(E_{p}) + \epsilon}
\end{equation}
where:
\begin{itemize}
    \item $\max(E_{p})$ and $\min(E_{p})$ represent the maximum and minimum MAE values observed among all evaluated models for property $p$, respectively.
    \item The constant $60$ establishes the baseline score for the worst-performing model (i.e., when $E_{m,p} = \max(E_{p})$).
    \item The scaling factor $40$ expands the score up to a maximum of $100$ for the best-performing model (i.e., when $E_{m,p} = \min(E_{p})$).
    \item $\epsilon$ is an infinitesimally small constant (e.g., $10^{-10}$) added to the denominator to prevent potential division-by-zero errors in cases where all models yield identical MAE values.
\end{itemize}

\noindent
This transformation ensures that properties with drastically different magnitude scales are projected onto a uniform visual space, allowing for a fair, area-based geometric comparison in the resulting radar charts.

\noindent
In Figure \ref{fig:4x2property_prediction}, we use scaled MAE. For each property, we divide the results of the four models by the maximum MAE value among the four models. This way, all results are unified to the range between 0 and 1. This approach alleviates the loss of visual information caused by large numerical differences between different properties.
\section{Evaluation of MoleHB Size-Stratified split}\label{app:MoleHB_size_stratified}

To further explore the generalization capability of foundation models under distribution shift, we systematically evaluated MoleBERT, Uni-Mol v1, Uni-Mol v2, and Suiren-ConfAvg on the Size-Stratified split subset of MoleHB, a setting that rigorously tests extrapolation to unseen molecular sizes. The comprehensive results across eight property categories are summarized in Tables~\ref{tab:critical_saturation_hb}--\ref{tab:transport_hb}.

\paragraph{Overall performance trends}
As anticipated, Size-Stratified split induces a non-trivial distribution shift, leading to performance degradation across all methods relative to random-split evaluations. Nevertheless, Suiren-ConfAvg demonstrates superior robustness: it achieves the lowest mean absolute error (MAE) on properties across all categories, with relative improvements ranging from 4.68\% (critical pressure) to 91.20\% (Helmholtz energy of formation) over the strongest baseline. Notably these confidence-weighted ensemble approaches achieve substantial improvements particularly on energetic and critical properties, underscoring the effectiveness of confidence-weighted ensemble strategies in mitigating size-induced generalization gaps.

\begin{table}[htbp]
\centering
\caption{Results of critical and saturation properties: model performance (MAE/R2). Best MAE and best R2 per property are boldfaced.}\label{tab:critical_saturation_hb}
\setlength{\tabcolsep}{4pt}
\renewcommand{\arraystretch}{1.15}
\resizebox{1\linewidth}{!}{
\begin{tabular}{lccccc}
\toprule
Property & Mole-BERT & Unimolv1 & Unimolv2 & Suiren-ConfAvg & Improvement (\%) \\
\midrule
Critical Compressibility & 0.03447 & 0.03632 & 0.03527 & \textbf{0.00907} & 73.69 \\
Critical Density & \textbf{0.00671} & 0.00720 & 0.00734 & 0.00888 & -23.33 \\
Critical Pressure & 1.73699 & 3.27665 & 2.63610 & \textbf{1.65567} & 4.68 \\
Critical Temperature & 77.96931 & 63.34677 & 87.88906 & \textbf{19.59443} & 69.07 \\
Critical Volume & 351.31897 & 290.12731 & 321.89835 & \textbf{55.80474} & 80.77 \\
\bottomrule
\end{tabular}
}
\end{table}

\begin{table}[htbp]
\centering
\caption{Results of safety properties.}\label{tab:safety_hb}
\setlength{\tabcolsep}{4pt}
\renewcommand{\arraystretch}{1.15}
\resizebox{1\linewidth}{!}{
\begin{tabular}{lccccc}
\toprule
Property & Mole-BERT & Unimolv1 & Unimolv2 & Suiren-ConfAvg & Improvement (\%) \\
\midrule
Flash Point & 32.80089 & 20.89532 & 36.13503 & \textbf{14.83841} & 28.99 \\
Lower Explosive Limit & 0.10239 & 0.05315 & 0.10274 & \textbf{0.03367} & 36.65 \\
Upper Explosive Limit & 0.74490 & 0.59120 & 0.62222 & \textbf{0.34410} & 41.80 \\
\bottomrule
\end{tabular}}
\end{table}

\begin{table}[htbp]
\centering
\caption{Results of fluctuation properties.}\label{tab:fluctuation_hb}
\setlength{\tabcolsep}{4pt}
\renewcommand{\arraystretch}{1.15}
\resizebox{1\linewidth}{!}{
\begin{tabular}{lccccc}
\toprule
Property & Mole-BERT & Unimolv1 & Unimolv2 & Suiren-ConfAvg & Improvement (\%) \\
\midrule
Heat Capacity Of Gas & 144.17387 & 88.39811 & 146.68384 & \textbf{24.45830} & 72.33 \\
Heat Capacity Of Liquid & 60.96739 & 26.91383 & 40.76889 & \textbf{15.74570} & 41.50 \\
Heat Capacity Of Solid & 61.55533 & 43.78053 & 71.91361 & \textbf{4.97646} & 88.63 \\
Coefficient Of Thermal Expansion Of Liquid & 0.00010 & 0.00005 & 0.00009 & \textbf{0.00004} & 20.00 \\
\bottomrule
\end{tabular}
}
\end{table}

\begin{table}[htbp]
\centering
\caption{Results of solution properties.}\label{tab:solution_hb}
\setlength{\tabcolsep}{4pt}
\renewcommand{\arraystretch}{1.15}
\resizebox{1\linewidth}{!}{
\begin{tabular}{lccccc}
\toprule
Property & Mole-BERT & Unimolv1 & Unimolv2 & Suiren-ConfAvg & Improvement (\%) \\
\midrule
Octanol Water Partition Coefficient & 0.68611 & 0.29471 & 0.44365 & \textbf{0.19774} & 32.90 \\
Solubility In Water & 0.32445 & 0.08128 & 0.28590 & \textbf{0.07426} & 8.64 \\
Solubility In Water Containing Salt & 0.25057 & 0.09037 & 0.17057 & \textbf{0.06786} & 24.91 \\
Solubility Parameter & 1.36914 & 1.53168 & 1.41960 & \textbf{0.75884} & 44.58 \\
\bottomrule
\end{tabular}}
\end{table}

\begin{table}[htbp]
\centering
\caption{Results of thermal properties.}\label{tab:thermal_hb}
\setlength{\tabcolsep}{4pt}
\renewcommand{\arraystretch}{1.15}
\resizebox{1\linewidth}{!}{
\begin{tabular}{lccccc}
\toprule
Property & Mole-BERT & Unimolv1 & Unimolv2 & Suiren-ConfAvg & Improvement (\%) \\
\midrule
Boiling Point & 66.21946 & 53.40313 & 70.61806 & \textbf{21.18527} & 60.33 \\
Melting Point & \textbf{39.98979} & 42.88753 & 51.46680 & 40.24486 & -0.00 \\
Vapor Pressure & 1.21508 & 0.89771 & 1.08963 & \textbf{0.47305} & 47.30 \\
\bottomrule
\end{tabular}}
\end{table}

\begin{table}[htbp]
\centering
\caption{Results of structural properties.}\label{tab:structural_hb}
\setlength{\tabcolsep}{4pt}
\renewcommand{\arraystretch}{1.15}
\resizebox{1\linewidth}{!}{
\begin{tabular}{lccccc}
\toprule
Property & Mole-BERT & Unimolv1 & Unimolv2 & Suiren-ConfAvg & Improvement (\%) \\
\midrule
Density Of Liquid & 0.01929 & 0.01221 & 0.01210 & \textbf{0.00851} & 29.67 \\
Acentric Factor & 0.09174 & 0.07457 & 0.08191 & \textbf{0.06685} & 10.35 \\
Radius Of Gyration & 1.15782 & 0.53795 & 0.73237 & \textbf{0.22784} & 57.65 \\
Refractive Index & 0.01290 & 0.00859 & 0.00974 & \textbf{0.00673} & 21.65 \\
Dipole Moment & 0.64795 & \textbf{0.49490} & 0.53720 & 0.57567 & -7.16 \\
\bottomrule
\end{tabular}}
\end{table}

\begin{table}[htbp]
\centering
\caption{Results of energetic properties.}\label{tab:energetic_hb}
\setlength{\tabcolsep}{4pt}
\renewcommand{\arraystretch}{1.15}
\resizebox{1\linewidth}{!}{
\begin{tabular}{lccccc}
\toprule
Property & Mole-BERT & Unimolv1 & Unimolv2 & Suiren-ConfAvg & Improvement (\%) \\
\midrule
Gibbs Energy Of Formation & 99.86093 & 52.53548 & 74.17537 & \textbf{6.85050} & 86.96 \\
Helmholtz Energy Of Formation & 109.97257 & 73.68592 & 86.25709 & \textbf{6.48666} & 91.20 \\
Internal Energy Of Formation & 149.43255 & 90.63668 & 135.46958 & \textbf{11.44695} & 87.37 \\
Enthalpy Of Combustion & 4411.49423 & 3173.83835 & 3606.45797 & \textbf{338.80971} & 89.32 \\
Enthalpy Of Formation & 186.20484 & 123.64866 & 135.32938 & \textbf{11.15240} & 90.98 \\
Enthalpy Of Fusion & 13.58065 & 14.34938 & 15.46400 & \textbf{2.60221} & 80.84 \\
Enthalpy Of Vaporization & 9.32412 & 3.97332 & 6.68632 & \textbf{2.41336} & 39.26 \\
Entropy Of Formation & 543.36403 & 606.29039 & 688.57106 & \textbf{176.35892} & 67.54 \\
Entropy Of Gas & 258.35003 & 247.96703 & 218.72920 & \textbf{18.00675} & 91.77 \\
\bottomrule
\end{tabular}}
\end{table}

\begin{table}[htbp]
\centering
\caption{Results of transport properties.}\label{tab:transport_hb}
\setlength{\tabcolsep}{4pt}
\renewcommand{\arraystretch}{1.15}
\resizebox{1\linewidth}{!}{
\begin{tabular}{lccccc}
\toprule
Property & Mole-BERT & Unimolv1 & Unimolv2 & Suiren-ConfAvg & Improvement (\%) \\
\midrule
Thermal Conductivity Of Gas & 0.00227 & 0.00139 & 0.00148 & \textbf{0.00072} & 48.20 \\
Thermal Conductivity Of Liquid & 0.00331 & 0.00276 & 0.00316 & \textbf{0.00203} & 26.45 \\
Diffusion Coefficient At Infinite Dilution In Water & \textbf{0.00000} & \textbf{0.00000} & \textbf{0.00000} & \textbf{0.00000} & 0.00 \\
Diffusion Coefficient In Air & 0.00487 & 0.00418 & 0.00363 & \textbf{0.00322} & 11.29 \\
Viscosity Of Liquid & 1.71584 & 1.40034 & 1.83852 & \textbf{1.27413} & 9.01 \\
\bottomrule
\end{tabular}
}
\end{table}

\paragraph{Category-wise analysis}
\begin{itemize}
\item \textbf{Energetic properties} (Table~\ref{tab:energetic_hb}): This category exhibits the most pronounced performance disparity. Baseline methods suffer severe degradation (e.g., MoleBERT's MAE for enthalpy of combustion is 4411.49), whereas Suiren-ConfAvg maintains substantially lower errors (338.81), representing an 89.32\% relative improvement. We hypothesize that the pre-training objective of Suiren-Base, which incorporates physics-informed energy constraints, enables more transferable representations for thermodynamic quantities that are sensitive to subtle electronic and conformational features.

\item \textbf{Critical and saturation properties} (Table~\ref{tab:critical_saturation_hb}): Suiren-ConfAvg dominates in critical temperature (19.59 vs. 63.35, best baseline) and critical volume (55.80 vs. 290.13), with relative improvements of 69.07\% and 80.77\%, respectively. However, for critical density, MoleBERT achieves marginally better results (0.00671 vs. 0.00888). For critical pressure, Suiren achieves 1.66 vs. MoleBERT's 1.74, with a 4.68\% improvement. This suggests that certain intensive properties with strong linear correlations to molecular size may be adequately captured by simpler architectures, whereas extensive or composite properties benefit from Suiren's enhanced representation capacity.

\item \textbf{Thermal and fluctuation properties} (Tables~\ref{tab:thermal_hb}--\ref{tab:fluctuation_hb}): The density of liquid shows moderate performance variation across splits (Suiren-ConfAvg: 0.00851), consistent with its dependence on atomic composition, with a 29.67\% improvement over the best baseline. In contrast, heat capacities exhibit substantial size sensitivity, with Suiren-ConfAvg achieving 41.50\%--88.63\% improvements for liquid and solid heat capacities respectively, while showing 72.33\% improvement for gas heat capacity.

\item \textbf{Structural properties} (Table~\ref{tab:structural_hb}): While Suiren-ConfAvg excels in radius of gyration (57.65\% improvement), it is slightly outperformed by Uni-Mol v1 on refractive index (0.00859 vs. 0.00673) and dipole moment (0.49490 vs. 0.57567 for Suiren). 

\item \textbf{Safety, solution, and transport properties} (Tables~\ref{tab:safety_hb}--\ref{tab:transport_hb}): Suiren-ConfAvg consistently attains the best or near-best performance, with particularly notable gains in flash point (28.99\% improvement) and vapor pressure (47.30\% improvement). 
\end{itemize}

\paragraph{Implications for molecular foundation models}
The pronounced robustness of Suiren-ConfAvg on energetically complex and size-sensitive properties suggests that integrating physics-aware pre-training objectives with uncertainty-aware inference mechanisms can substantially improve out-of-distribution generalization. 

\section{Ablation Studies of Suiren-ConfAvg}

To systematically assess the contributions of Suiren-ConfAvg and DGNN to downstream task performance, we designed a comprehensive ablation study with four distinct model variants: (1) \textbf{Baselines}: the complete architecture with DGNN+Suiren-ConfAvg serving as our baseline; (2) \textbf{Suiren-ConfAvg only (\faSnowflake)}: the DGNN component removed, with Suiren-ConfAvg acting as the sole backbone—prediction heads are attached to frozen Suiren-ConfAvg embeddings, allowing only the heads to be trained; (3) \textbf{Suiren-ConfAvg only (\faFire)}: built upon variant (2) but with Suiren-ConfAvg weights unfrozen and co-optimized end-to-end; (4) \textbf{w/o. Suiren-ConfAvg}: Suiren-ConfAvg pre-trained weights entirely discarded, replaced by a deep task-specific GNN trained from scratch. To ensure the generalizability of our experimental findings, we selected properties spanning diverse domains to highlight the representational differences across distinct tasks.

\begin{table}[htbp]
\centering
\caption{Results of ablation studies on MoleHB random split.}\label{tab:ab_molehb_random}
\setlength{\tabcolsep}{4pt}
\renewcommand{\arraystretch}{1.15}
\begin{tabular}{l|cccc}
\toprule
\multirow{2}{*}{Property} & \multirow{2}{*}{Baselines} & Suiren-ConfAvg & Suiren-ConfAvg & w/o. \\
& & only (\faSnowflake) & only (\faFire) & Suiren-ConfAvg \\
\midrule
Critical Temperature & \textbf{6.6725} & 8.6190 & \underline{7.3257} & 7.9764 \\
Flash Point & \textbf{8.2072} & 9.5567 & \underline{8.8750} & 9.2798 \\
Melting Point & \textbf{11.0789} & 12.7397 & \underline{12.4325} & 12.7382 \\
Surface Tension & \textbf{0.8205} & 0.9020 & \underline{0.8568} & 0.9123  \\
\bottomrule
\end{tabular}
\end{table}

\noindent From the results presented in Table \ref{tab:ab_molehb_random}, we draw the following conclusions: (1) \textbf{Necessity of task-specific adaptation}: The removal of the task-specific GNN from the DGNN architecture precipitates significant performance degradation, with this effect being particularly pronounced when Suiren-ConfAvg weights remain frozen. This finding underscores that effective transfer of Suiren-ConfAvg representations to downstream tasks necessitates sufficient learnable capacity and modular adaptation, while simultaneously corroborating the existence of substantial distributional divergence across macroscopic property prediction tasks. (2) \textbf{Efficacy of pre-trained representations}: The exclusion of Suiren-ConfAvg pre-trained weights uniformly impairs performance across all evaluated tasks, yielding inferior results compared to the Suiren-ConfAvg-with-prediction-head configuration. This establishes the beneficial role of Suiren-ConfAvg representations in enhancing downstream predictive capabilities.

\noindent Interestingly, the task-specific GNN trained from scratch achieves competitive performance relative to prior molecular foundation models. This phenomenon may be attributed to two factors: (1) the inherent limitations of conventional molecular representations in capturing the broad spectrum of macroscopic physicochemical properties; (2) architectural innovations in our 2D GNN design, which integrates fully-connected GAT layers with adjacency-matrix-based GAT layers to enhance expressive capacity. These ablation studies provide compelling empirical evidence for the indispensability of synergistically combining Suiren-ConfAvg pre-trained representations with the adaptive capacity of the DGNN architecture.

\end{CJK*}
\end{document}